\documentclass[twocolumn]{revtex4}
\usepackage{makeidx}
\usepackage{mathbbold}
\usepackage{amsfonts}
\usepackage{amsmath}
\usepackage{amssymb}
\usepackage{charter}
\usepackage{graphicx}
\usepackage{subfigure}

\begin{document}

\title{Coupled three-mode squeezed vacuum: Gaussian steering and remote
generation of Wigner negativity}
\author{Zi-wei Zhan, Bo Lan, Jian Wang and Xue-xiang Xu$^{\dag }$}
\affiliation{College of Physics and Communication Electronics, Jiangxi Normal University,
Nanchang 330022, China;\\
$^{\dag}$xuxuexiang@jxnu.edu.cn}

\begin{abstract}
Multipartite Einstein-Podolsky-Rosen (EPR) steering and multimode quantum
squeezing are essential resources for various quantum applications. The
paper focuses on studying a coupled three-mode squeezed vacuum (C3MSV),
which is a typical multimode squeezed Gaussian state and will exhibit
peculiar steering property. Using the technique of integration within
ordered products, we give the normal-ordering form for the coupled
three-mode squeezing operator and derive the general analytical expressions
of the statistical quantities for the C3MSV. Under Gaussian measurements, we
analyze all bipartite Gaussian steerings (including no steering, one-way
steering and two-way steering) in details and study the monogamy relations
for the C3MSV. Then, we study the decoherence of all these steerings in
noisy channels and find that sudden death will happen in a certain threshold
time. Through the steerings shared in the C3MSV, we propose conceptual (and
ideal) schemes of remotely generating Wigner negativity (WN) by performing
appropriate photon subtraction(s) in the local position. Our obtained
results may lay a solid theoretical foundation for a future practical study.
We also believe that the C3MSV will be one of good candidate resources in
future quantum protocols.

\end{abstract}

\maketitle

\section{Introduction}

Quantum correlations have been intensively investigated in recent years and
can manifest in different forms, such as entanglement, steering and Bell
nonlocality \cite{1}. These correlations are established in two-party,
three-party, or even more-party systems and can be used as resources for
quantum enhanced tasks \cite{2,3,4,5}. Entanglement is a striking feature of
describing the nonbiseparability of states for two or more parties \cite{6}.
Bell nonlocality offers a vast research landscape with relevance for
fundamental \cite{7} and quantum technological applications \cite{8,9}.
Eistein-Podolsky-Rosen (EPR) steering is intermediate between entanglement
and Bell nonlocality \cite{10,11,12,13,14}. Its concept\ was named by
Schrodinger \cite{15} and rigorously defined by Wiseman \textit{et al.} \cite%
{16,17}.\ EPR steering\ is often the required resource enabling the protocol
to proceed securely \cite{18} and has been applied to realize different
tasks \cite{19}.

Over the past several decades, significant advances on squeezed light
generation have been made \cite{20,21}.\ Squeezed optical fields,
particularly those states with multimode squeezing, are essential resources
in quantum technologies \cite{22}. Nonlinear optics provides a number of
promising experimental tools for realizing multipartite correlation and
multimode squeezing \cite{23}. One conventional tool is to employ the
optical parametric oscillator technique \cite{24}. Another mature tool is to
employ a four-wave mixing (FWM) process \cite{25}. FWM describes a
parametric interaction between four coherent fields in a nonlinear crystal
\cite{26}.

There is a tendency for researchers to use multipartite quantum correlations
and multimode quantum squeezing as resources. Specially, EPR steering in a
multipartite scenario has been used for the implementation of secure
multiuser quantum technologies \cite{27}. Many schemes of generating
multimode squeezed and correlated states have been proposed.\ Their common
kernel idea is based on the basic FWM process by using multiple pump beams%
\cite{28}, spatially structured pump beams \cite{29,30,31} or cascading
setups \cite{32,33,34}. These schemes of cascaded FWM processes can be used
to generate \cite{35,36,37} and even enhance \cite{38} multipartite
entanglement.

A two-mode squeezed vacuum (TMSV) is perhaps the most commonly used EPR
entangled resource\cite{3}. Rather than a TMSV, many entangled resources
(such as the NOON state \cite{39,40} and the Greenberger-Horne-Zeilinger
state \cite{41}) have been also used in other scenarios. With the
development and requirements of quantum technology, more and more entangled
resources have been introduced and used\cite{42,43,44,45,46}.\ Based on the
energy-level cascaded FWM system, Qin \textit{et al.} constructed 11
possible Hamiltonians, which may help to generate three-mode and four-mode
quantum squeezed states\cite{47}. Qin and co-worker generated triple-beam
quantum-correlated states, which may show the tripartite entanglement\cite%
{34}. By FWM with linear and nonlinear beamsplitters, Liu \textit{et al.}
introduced a three-mode Gaussian state\cite{48}, which may exhibit
tripartite EPR steering. Li \textit{et al.} also generated
quantum-correlated three-mode light beams\cite{49}. Zhang and Glasser\cite%
{50} introduced a coupled three-mode squeezed vacuum (C3MSV), which exhibits
genuinely tripartite entanglement.

On the other hand, Wigner negativity (WN) \cite{51} is arguably one of the
most striking non-classical features of quantum states and has been
attracting increasing interests \cite{52}. Beyond its fundamental relevance
\cite{53,54}, WN is also a necessary resource for quantum speedup with
continuous variables. It has been seen as a necessary ingredient in
continuous-variable quantum computation and simulation to outperform
classical devices \cite{55,56}. As two important signatures of
nonclassicality, quantum correlations can be intertwined with WN in the
conditional generation of non-Gaussian states \cite{57,58}. Walschaers
\textit{et al.} developed a general formalism to prepare Wigner-negative
states through EPR steering \cite{59,60,61}. Xiang \textit{et al.} proposed
schemes for remote generation of WN through EPR steering in a multipartite
scenario \cite{62}, where they used a pure three-mode Gaussian state
(realized by a feasible linear optical network) as the resource.

Intuitively, we think that the C3MSV will become an useful entangled
resource in future quantum protocols. Except those properties such as
squeezing and entanglement considered by Zhang and Glasser \cite{50}, we
will further study steering properties for the C3MSV in this paper.
Considering the effect of the environment, we also study the decoherence of
the steering. And then, we will propose schemes of remote preparation of
Wigner negative states. One can refer to the appendixes for the derivation
results and to the Supplemental Material \cite{63} for the codes. The rest
of the paper is structured as follows: In Sec.II, we make a brief
introduction of the coupled three-mode squeezing operator (C3MSO) and the
C3MSV. In Sec.III, we investigate the bipartite Gaussian steerings in the
C3MSV. In Sec.IV, we study the decoherence of the steering. In Sec.V, we
propose schemes to remotely generate WN based on the steering in the C3MSV.
Conclusions are summarized in the last section.

\section{Coupled three-mode squeezed vacuum}

An interaction with the three-mode Hamiltonian\ $H_{I}=i\hbar (\eta
_{1}^{\ast }a_{1}a_{2}$ $+\eta _{2}^{\ast }a_{2}a_{3}$ $-\eta
_{1}a_{1}^{\dagger }a_{2}^{\dagger }$ $-\eta _{2}a_{2}^{\dagger
}a_{3}^{\dagger })$\ can be realized by using a dual-pumping FWM process,
where $a_{j}$ ($a_{j}^{\dagger }$)\ is the bosonic annihilation (creation)
operator in mode $j$. The detailed description of the interaction has been
provided by Zhang and Glasser \cite{50}. Associated with this Hamiltonian,
one can obtain the following unitary time evolution operator (i.e.,\ the
C3MSO)%
\begin{equation}
S_{3}=e^{\xi _{1}^{\ast }a_{1}a_{2}+\xi _{2}^{\ast }a_{2}a_{3}-\xi
_{1}a_{1}^{\dagger }a_{2}^{\dagger }-\xi _{2}a_{2}^{\dagger }a_{3}^{\dagger
}},  \label{1-1}
\end{equation}%
where $\xi _{j}=\eta _{j}t=r_{j}e^{i\theta _{j}}$ ($j=1$ and $2$) are the
two complex squeezing parameters, with respective magnitude $r_{j}$\ and
phase $\theta _{j}$. It is obvious to see $S_{3}^{-1}=S_{3}^{\dag }$. For
convenience, we reset ($r_{1}$, $r_{2}$) as ($r$, $\phi $), satisfying $r=%
\sqrt{r_{1}^{2}+r_{2}^{2}}$, $\cos \phi =r_{1}/r$, and $\sin \phi =r_{2}/r$
with $\phi \in \lbrack 0,\pi /2]$ [see Fig.1(a)]. A similar three-mode
squeezing interaction has also been analyzed theoretically and realized
experimentally by Paris's group. By interlinked nonlinear interactions in $%
\chi ^{(2)}$ media, they addressed the generation of fully inseparable
three-mode entangled states of radiation \cite{64,65}. In addition, they
applied this three-mode entanglement in realizing symmetric and asymmetric
telecloning machines and generalized these studies to multimode cases \cite%
{66}.

As illustrated in Fig.1(b), by applying the C3MSO $S_{3}$ on the three
independent vacuum $\left\vert 0\right\rangle \left\vert 0\right\rangle
\left\vert 0\right\rangle $, we easily obtain the C3MSV with the following
form%
\begin{equation}
\left\vert \psi \right\rangle \equiv S_{3}\left\vert 000\right\rangle =\frac{%
1}{c}e^{-\frac{\epsilon _{1}}{c}a_{1}^{\dag }a_{2}^{\dag }-\frac{\epsilon
_{2}}{c}a_{2}^{\dag }a_{3}^{\dag }}\left\vert 000\right\rangle ,  \label{1-2}
\end{equation}%
whose density operator is $\rho _{123}=\left\vert \psi \right\rangle
\left\langle \psi \right\vert $. In Appendix A, we have\ given the
normal-ordering form for the C3MSO by using the technique of integration
within ordered products (IWOP) \cite{67,68}. Here, we set $c=\cosh r$, $%
s=\sinh r$, $\epsilon _{1}=se^{i\theta _{1}}\cos \phi $, and $\epsilon
_{2}=se^{i\theta _{2}}\sin \phi $. In particular, if $\xi _{2}=0,$ then $%
\left\vert \psi \right\rangle =S_{2}(\xi _{1})\left\vert 00\right\rangle
_{12}\otimes \left\vert 0\right\rangle _{3}$; if $\xi _{1}=0,$ then $%
\left\vert \psi \right\rangle =\left\vert 0\right\rangle _{1}\otimes
S_{2}(\xi _{2})\left\vert 00\right\rangle _{23}$, with $S_{2}(\xi
_{1})=e^{\xi _{1}^{\ast }a_{1}a_{2}-\xi _{1}a_{1}^{\dagger }a_{2}^{\dagger
}} $\ and $S_{2}(\xi _{2})=e^{\xi _{2}^{\ast }a_{2}a_{3}-\xi
_{2}a_{2}^{\dagger }a_{3}^{\dagger }}$. Moreover, if $\xi _{1}=\xi _{2}$
(i.e. $\phi =\pi /4$), the C3MSV is a bisymmetric state, whose mode 1 and
mode 3 are symmetrical with mode 2. Zhang and Glasser have analyzed the
squeezing property and the entanglement characteristics for the C3MSV\cite%
{50}, which further\ reflect that the C3MSO has the utility of realizing
available squeezing and genuine tripartite entanglement.

Using\ the general expression for the C3MSV in Eq.(A4), we easily obtain $%
\bar{n}_{1}=s^{2}\cos ^{2}\phi $, $\bar{n}_{2}=s^{2}$, $\bar{n}%
_{3}=s^{2}\sin ^{2}\phi $, and $\bar{n}_{T}=2s^{2}$, i.e., the mean photon
numbers (MPNs) for mode 1, mode 2, mode 3 and total modes, respectively [see
Fig.1(c)]. By the way, we often replace $r$ by $\bar{n}_{T}$ (using $r=$%
arcsinh$\sqrt{\bar{n}_{T}/2}$) and set $\theta _{1}=\theta _{2}=0$ in our\
following numerical work.
\begin{figure}[tbp]
\label{Fig1} {\centering
\includegraphics[width=0.95\columnwidth]{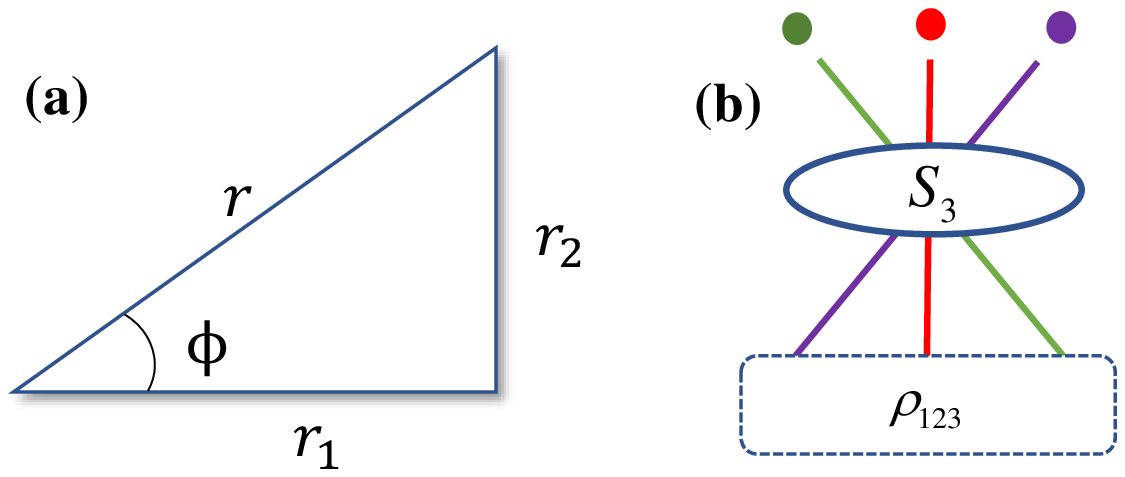}} {\centering
\includegraphics[width=0.8\columnwidth]{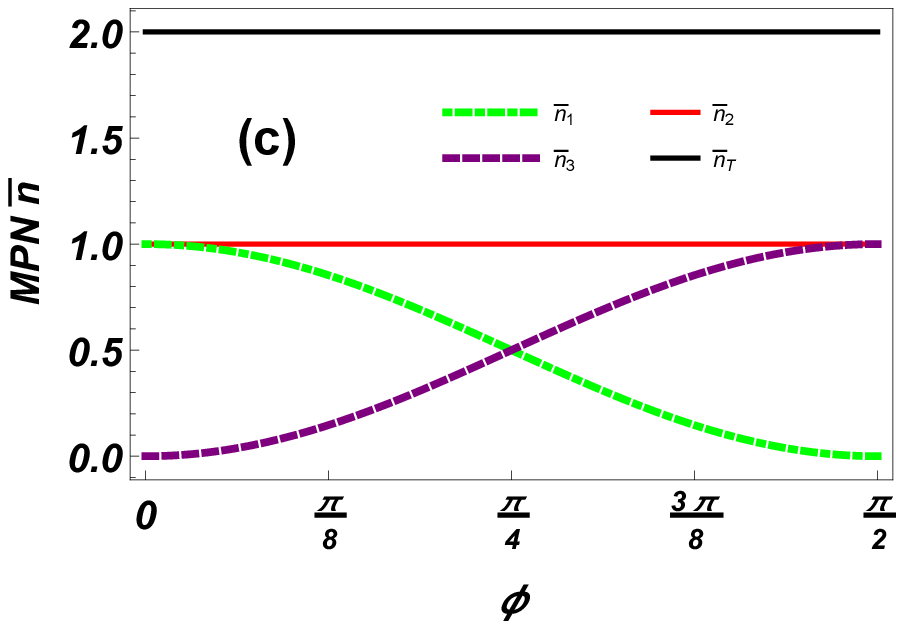}}
\caption{(a) ($r_{1}$, $r_{2}$) are reset to ($r$, $\protect\phi $). (b)
Conceptual generating scheme of the C3MSV $\protect\rho _{123}$, which is
obtained by applying $S_{3}$\ on three independent vacuums $\left\vert
000\right\rangle $ (represented by three small balls separately at the top).
(c) MPNs $\bar{n}_{1}$, $\bar{n}_{2}$, $\bar{n}_{3}$, and $\bar{n}_{T}$
versus $\protect\phi $ (setting $\bar{n}_{T}=2$).}
\end{figure}

\section{Gaussian steering in the C3MSV}

The C3MSV is a pure three-mode entangled Gaussian state, which can be seen
from its Wigner function provided in Eq. (C1). In this section, we analyze
the distributions of bipartite\ Gaussian steerings in the C3MSV, without
considering the optical losses and thermal noises.

\subsection{Covariance matrix of the C3MSV}

The covariance matrix (CM) \cite{69,70,71,72,73} of the C3MSV is expressed as

\begin{equation}
V=\left(
\begin{array}{ccc}
(1+2\bar{n}_{1})\text{ }1_{2} & -2sc\cos \phi \text{ }\Sigma _{\theta _{1}}
& s^{2}\sin 2\phi \text{ }R_{\theta _{2}-\theta _{1}} \\
-2sc\cos \phi \text{ }\Sigma _{\theta _{1}} & (1+2\bar{n}_{2})\text{ }1_{2}
& -2sc\sin \phi \text{ }\Sigma _{\theta _{2}} \\
s^{2}\sin 2\phi \text{ }\tilde{R}_{\theta _{2}-\theta _{1}} & -2sc\sin \phi
\text{ }\Sigma _{\theta _{2}} & (1+2\bar{n}_{3})\text{ }1_{2}%
\end{array}%
\right) ,  \label{2-1}
\end{equation}%
with the $2\times 2$ identity matrix $1_{2}$ and%
\begin{equation}
\Sigma _{\theta }=\left(
\begin{array}{cc}
\cos \theta & \sin \theta \\
\sin \theta & -\cos \theta%
\end{array}%
\right) ,R_{\theta }=\left(
\begin{array}{cc}
\cos \theta & \sin \theta \\
-\sin \theta & \cos \theta%
\end{array}%
\right) .  \label{2-2}
\end{equation}%
The matrix elements of the CM, defined by $V_{jk}=\left\langle \psi
\right\vert (\hat{X}_{j}\hat{X}_{k}+\hat{X}_{k}\hat{X}_{j})\left\vert \psi
\right\rangle $, are expressed via the vector $\hat{X}=(\hat{x}_{1},\hat{p}%
_{1},\hat{x}_{2},\hat{p}_{2},\hat{x}_{3},\hat{p}_{3}$). For each mode, we
define the position operator $\hat{x}_{j}=\frac{1}{\sqrt{2}}(\hat{a}_{j}+%
\hat{a}_{j}^{\dag })$\ and the momentum operator $\hat{p}_{j}=\frac{1}{i%
\sqrt{2}}(\hat{a}_{j}-\hat{a}_{j}^{\dag })$,\ accompanied by its
annihilation and creation operators $\hat{a}_{j}$ and $\hat{a}_{j}^{\dag }$.
It is noted that $\left\langle \hat{x}_{j}\right\rangle =\left\langle \hat{p}%
_{j}\right\rangle =0$ for each mode of the C3MSV. The CM $V$ in Eq.(\ref{2-1}%
) is a symmetric and positive semidefinite matrix (with eigenvalues $1$, $1$%
, $e^{-2r}$, $e^{-2r}$, $e^{2r}$, and $e^{2r}$) and obeys $V+i\Omega
^{\oplus 3}\geq 0$ with $\Omega =\left(
\begin{array}{cc}
0 & 1 \\
-1 & 0%
\end{array}%
\right) $. Moreover, we can check $\det V=1$ and prove that the C3MSV is a
pure state.

\subsection{Bipartite Gaussian steering}

Quantum protocols often require only states (e.g., squeezed vacuum states)
and measurements (e.g., homodyne detection) that are simple to realize on
quantum optics platforms. Undoubtedly, the C3MSV is a good candidate
Gaussian state. Meanwhile, one can explore the Gaussian steerings by
Gaussian measurements \cite{74,75}. Moreover, the distribution of the
steering can be constrained by its monogamy relation. Reid derived monogamy
inequalities for the bipartite EPR steering distributed among different
systems \cite{76}. Xiang \textit{et al.} derived the laws for the
distribution of quantum steering among different parties and proved a
monogamy relation of Gaussian steering \cite{77}.

The CM of a bipartite Gaussian state can be expressed as%
\begin{equation}
\sigma _{AB}=\left(
\begin{array}{cc}
V_{A} & V_{AB} \\
V_{AB}^{T} & V_{B}%
\end{array}%
\right) ,  \label{2-3}
\end{equation}%
where party $A$ and party $B$\ are the bipartite subsystems. Then, we can
quantify how much it is steerable via the following quantity%
\begin{equation}
\mathcal{G}^{A\rightarrow B}\left( V\right) :=\max \{0,-\sum_{j:\text{ }\bar{%
\nu}_{j}^{B|A}<1}\ln \bar{\nu}_{j}^{B|A}\}.  \label{2-4}
\end{equation}%
where $\{\bar{\nu}_{j}^{B|A}\}_{j=1}^{2n_{B}}$ denote the symplectic
eigenvalues ($n_{B}$ is the mode number in subsystem $B$) of the Schur
complement $\sigma _{B|A}=V_{B}-V_{AB}^{T}V_{A}^{-1}V_{AB}$ of $\sigma _{AB}$%
. Obviously, the mathematical formalism of Gaussian steering $\mathcal{G}%
^{A\rightarrow B}$\ is achieved by Gaussian measurements in party $A$. This
quantity $\mathcal{G}^{A\rightarrow B}$ is defined as Gaussian $A\rightarrow
B$ steerability, which is a monotone under Gaussian local operations and
classical communication. Moreover, the larger $\mathcal{G}^{A\rightarrow B}$
is, the stronger Gaussian steerability is \cite{10,78}.
\begin{figure}[tbp]
\label{Fig2} \centering\includegraphics[width=1.0\columnwidth]{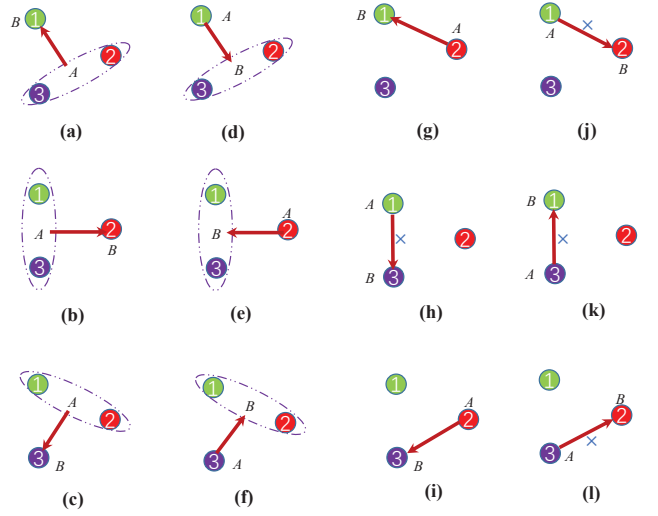}
\caption{Bipartite assignments in the C3MSV, where $A$ is the steering party
and $B$ is the steered party.}
\end{figure}

In what follows, we take party $A$ and party $B$ from the three modes of the
C3MSV and construct 12 kinds of $\sigma _{AB}$s from $V$ in Eq. (\ref{2-1}).
The steering party $A$ and the steered party $B$ are assigned as shown in
Fig.2. According to the rule in Eq. (\ref{2-4}), we obtain the following
steerings present in the C3MSV.

\textit{Case (a)}$\Longrightarrow $\textit{A(23)-B(1). }In this case, we
have $\bar{\nu}_{1}^{B|A}=\bar{\nu}_{2}^{B|A}=(c^{2}+s^{2}\cos 2\phi
)^{-1}<1 $, which leads to
\begin{equation}
\mathcal{G}^{23\rightarrow 1}=2\ln (c^{2}+s^{2}\cos 2\phi ).  \label{2-5}
\end{equation}

\textit{Case (b)}$\Longrightarrow $\textit{A(13)-B(2).} In this case, we
have $\bar{\nu}_{1}^{B|A}=\bar{\nu}_{2}^{B|A}=(c^{2}+s^{2})^{-1}<1$, which
leads to
\begin{equation}
\mathcal{G}^{13\rightarrow 2}=2\ln (c^{2}+s^{2}).  \label{2-6}
\end{equation}

\textit{Case (c)}$\Longrightarrow $\textit{A(12)-B(3). }In this case, we
have $\bar{\nu}_{1}^{B|A}=\bar{\nu}_{2}^{B|A}=(c^{2}-s^{2}\cos 2\phi
)^{-1}<1 $, which leads to
\begin{equation}
\mathcal{G}^{12\rightarrow 3}=2\ln (c^{2}-s^{2}\cos 2\phi ).  \label{2-7}
\end{equation}

\textit{Case (d)}$\Longrightarrow $\textit{A(1)-B(23).} In this case, we
have $\bar{\nu}_{1}^{B|A}=\bar{\nu}_{2}^{B|A}=\left( \varkappa _{1}-\sqrt{%
\varkappa _{2}}\right) /\varkappa _{0}<1$ and $\bar{\nu}_{3}^{B|A}=\bar{\nu}%
_{4}^{B|A}=\left( \varkappa _{1}+\sqrt{\varkappa _{2}}\right) /\varkappa
_{0}\geq 1$, which leads to
\begin{equation}
\mathcal{G}^{1\rightarrow 23}=2\ln [\varkappa _{0}/\left( \varkappa _{1}-%
\sqrt{\varkappa _{2}}\right) ],  \label{2-8}
\end{equation}%
with%
\begin{eqnarray}
\varkappa _{0} &=&4+8s^{2}\cos ^{2}\phi ,  \notag \\
\varkappa _{1} &=&1+3c^{2}+\left( 3-2\cos 2\phi \right) s^{2},  \notag \\
\varkappa _{2} &=&\left( 19-12\cos 2\phi \right) c^{2}s^{2}  \notag \\
&&+\left( 19-12\cos 2\phi +2\cos 4\phi \right) s^{4}  \notag \\
&&+\left( 13-20\cos 2\phi \right) s^{2}.  \label{2-9}
\end{eqnarray}

\textit{Case (e)}$\Longrightarrow $\textit{A(2)-B(13).} In this case, we
have $\bar{\nu}_{1}^{B|A}=\bar{\nu}_{2}^{B|A}=(c^{2}+s^{2})^{-1}<1$ and $%
\bar{\nu}_{3}^{B|A}=\bar{\nu}_{4}^{B|A}=1$, which leads to
\begin{equation}
\mathcal{G}^{2\rightarrow 13}=2\ln (c^{2}+s^{2}).  \label{2-10}
\end{equation}

\textit{Case (f)}$\Longrightarrow $\textit{A(3)-B(12).} In this case, we
have $\bar{\nu}_{1}^{B|A}=\bar{\nu}_{2}^{B|A}=\left( \iota _{1}-\sqrt{\iota
_{2}}\right) /\iota _{0}<1$ and $\bar{\nu}_{3}^{B|A}=\bar{\nu}%
_{4}^{B|A}=\left( \iota _{1}+\sqrt{\iota _{2}}\right) /\iota _{0}\geq 1$,
which leads to
\begin{equation}
\mathcal{G}^{3\rightarrow 12}=2\ln [\iota _{0}/\left( \iota _{1}-\sqrt{\iota
_{2}}\right) ],  \label{2-11}
\end{equation}%
with

\begin{eqnarray}
\iota _{0} &=&4+8s^{2}\sin ^{2}\phi ,  \notag \\
\iota _{1} &=&1+3c^{2}+(3+2\cos 2\phi )s^{2},  \notag \\
\iota _{2} &=&\left( 19+12\cos 2\phi \right) c^{2}s^{2}+  \notag \\
&&\left( 19+12\cos 2\phi +2\cos 4\phi \right) s^{4}  \notag \\
&&+\left( 13+20\cos 2\phi \right) s^{2}.  \label{2-12}
\end{eqnarray}

\textit{Case (g)}$\Longrightarrow $\textit{A(2)-B(1).} In this case, we have
$\bar{\nu}_{1}^{B|A}=\bar{\nu}_{2}^{B|A}=(c^{2}-s^{2}\cos 2\phi
)/(c^{2}+s^{2})<1$, which leads to
\begin{equation}
\mathcal{G}^{2\rightarrow 1}=2\ln [(c^{2}+s^{2})/(c^{2}-s^{2}\cos 2\phi )].
\label{2-13}
\end{equation}

\textit{Case (h)}$\Longrightarrow $\textit{A(1)-B(3).} In this case, we have
$\bar{\nu}_{1}^{B|A}=\bar{\nu}_{2}^{B|A}=(c^{2}+s^{2})/(1+2s^{2}\cos
^{2}\phi )\geq 1$,, which leads to
\begin{equation}
\mathcal{G}^{1\rightarrow 3}=0.  \label{2-14}
\end{equation}

\textit{Case (i)}$\Longrightarrow $\textit{A(2)-B(3).} In this case, we have
$\bar{\nu}_{1}^{B|A}=\bar{\nu}_{2}^{B|A}=(c^{2}+s^{2}\cos 2\phi
)/(c^{2}+s^{2})<1$, which leads to
\begin{equation}
\mathcal{G}^{2\rightarrow 3}=2\ln [(c^{2}+s^{2})/(c^{2}+s^{2}\cos 2\phi )].
\label{2-15}
\end{equation}

\textit{Case (j)}$\Longrightarrow $\textit{A(1)-B(2).} In this case, we have
$\bar{\nu}_{1}^{B|A}=\bar{\nu}_{2}^{B|A}=(c^{2}-s^{2}\cos 2\phi
)/(1+2s^{2}\cos ^{2}\phi )\geq 1$, which leads to
\begin{equation}
\mathcal{G}^{1\rightarrow 2}=0.  \label{2-16}
\end{equation}

\textit{Case (k)}$\Longrightarrow $\textit{A(3)-B(1).} In this case, we have
$\bar{\nu}_{1}^{B|A}=\bar{\nu}_{2}^{B|A}=(c^{2}+s^{2})/(1+2s^{2}\sin
^{2}\phi )\geq 1$, which leads to
\begin{equation}
\mathcal{G}^{3\rightarrow 1}=0.  \label{2-17}
\end{equation}

\textit{Case (l)}$\Longrightarrow $\textit{A(3)-B(2).} In this case, we have
$\bar{\nu}_{1}^{B|A}=\bar{\nu}_{2}^{B|A}=(c^{2}+s^{2}\cos 2\phi
)/(1+2s^{2}\sin ^{2}\phi )\geq 1$, which leads to
\begin{equation}
\mathcal{G}^{3\rightarrow 2}=0.  \label{2-18}
\end{equation}

More interestingly, all the above steerings are independent of phases\textbf{%
\ (}$\theta _{1}$\textbf{, }$\theta _{2}$\textbf{).} As we all know,\textit{%
\ }EPR steering\ is a directional form of nonlocality and possesses an
asymmetric property. This characteristic can also be reflected in our
steerings. In Fig.3, we draw the contour plots of $\mathcal{G}%
^{12\rightarrow 3}$, $\mathcal{G}^{1\rightarrow 23}$, $\mathcal{G}%
^{23\rightarrow 1}$, $\mathcal{G}^{3\rightarrow 12}$, $\mathcal{G}%
^{2\rightarrow 1}$, and $\mathcal{G}^{2\rightarrow 3}$\ in the ($\bar{n}_{T}$%
, $\phi $) space. Interestingly, $\mathcal{G}^{12\rightarrow 3}$\ with%
\textbf{\ }$\mathcal{G}^{1\rightarrow 23}$, $\mathcal{G}^{23\rightarrow 1}$\
with $\mathcal{G}^{3\rightarrow 12}$,\ and $\mathcal{G}^{2\rightarrow 1}$\
with $\mathcal{G}^{2\rightarrow 3}$, all have the symmetry\ on the axis $%
\phi =\pi /4$\textbf{. }This is due to self-characteristics of the C3MSV. In
Fig.4, we plot $\mathcal{G}^{A\rightarrow B}$s versus $\phi $ (with $\bar{n}%
_{T}=3$) and $\mathcal{G}^{A\rightarrow B}$s versus $\bar{n}_{T}$ (with $%
\phi =\pi /8$). Among them, $\mathcal{G}^{13\rightarrow 2}=\mathcal{G}%
^{2\rightarrow 13}$ are functions of $\bar{n}_{T}$ and are not independent
of $\phi $. Moreover, we know that $\mathcal{G}^{1\rightarrow 2}=\mathcal{G}%
^{1\rightarrow 3}=\mathcal{G}^{3\rightarrow 1}=\mathcal{G}^{3\rightarrow
2}=0 $, but $\mathcal{G}^{2\rightarrow 1}>0$ (except $\phi =\pi /2$), $%
\mathcal{G}^{2\rightarrow 3}>0$ (except $\phi =0$), and $\mathcal{G}%
^{2\rightarrow 1}=\mathcal{G}^{2\rightarrow 3}$ for $\phi =\pi /4$. As $\bar{%
n}_{T}$\ increasing, most of the $\mathcal{G}^{A\rightarrow B}$s will
increase. The results tell us that three types of steerings, i.e., no
steering ($A$ cannot steer $B$ and $B$ cannot steer $A$), one-way steering ($%
A$ can steer $B$ while $B$ cannot steer $A$), or two-way (symmetrical or
asymmetrical) steering ($A$ can steer $B$ and $B$ can steer $A$), are
presented in the C3MSV. The main results are summarized as follows.

(i) There is no steering between mode 1 and mode 3 because of $\mathcal{G}%
^{1\rightarrow 3}=0$\ and $\mathcal{G}^{3\rightarrow 1}=0$ [see Eq.(\ref%
{2-14}) and Eq.(\ref{2-17})].

(ii) There is one-way steering between mode 1 and mode 2 because of $%
\mathcal{G}^{2\rightarrow 1}>0$\ and $\mathcal{G}^{1\rightarrow 2}=0$ [see
Eq.(\ref{2-13}) and Eq.(\ref{2-16})].

(iii) There is one-way steering between mode 2 and mode 3 because of $%
\mathcal{G}^{2\rightarrow 3}>0$\ and $\mathcal{G}^{3\rightarrow 2}=0$ [see
Eq.(\ref{2-15}) and Eq.(\ref{2-18})].

(iv) There is two-way asymmetrical steering between mode 1 and group (23)
because of $\mathcal{G}^{23\rightarrow 1}>0$\ and $\mathcal{G}^{1\rightarrow
23}>0$ but $\mathcal{G}^{23\rightarrow 1}\neq \mathcal{G}^{1\rightarrow 23}$
[see Eq.(\ref{2-5}) and Eq.(\ref{2-8})].

(v) There is two-way symmetrical steering between mode 2 and group (13)
because of $\mathcal{G}^{13\rightarrow 2}=$ $\mathcal{G}^{2\rightarrow 13}>0$
[see Eq.(\ref{2-6}) and Eq.(\ref{2-10})].

(vi) There is two-way asymmetrical steering between mode 3 and group (12)
because of $\mathcal{G}^{12\rightarrow 3}>0$\ and $\mathcal{G}^{3\rightarrow
12}>0$ but $\mathcal{G}^{12\rightarrow 3}\neq \mathcal{G}^{3\rightarrow 12}$
[see Eq.(\ref{2-7}) and Eq.(\ref{2-11})].

Just like what He \textit{et al.} said in their work\cite{14}, our results
also show that each mode can be steered by one or both of the other two in
the C3MSV. Moreover, we find that (a) $\mathcal{G}^{2\rightarrow 1}>0$, but $%
\mathcal{G}^{3\rightarrow 1}=0$, and (b) $\mathcal{G}^{2\rightarrow 3}>0$,
but $\mathcal{G}^{1\rightarrow 3}=0$. This result holds the character that
two parties cannot steer the same system \cite{76}.
\begin{figure}[tbp]
\label{Fig3} \centering\includegraphics[width=0.9\columnwidth]{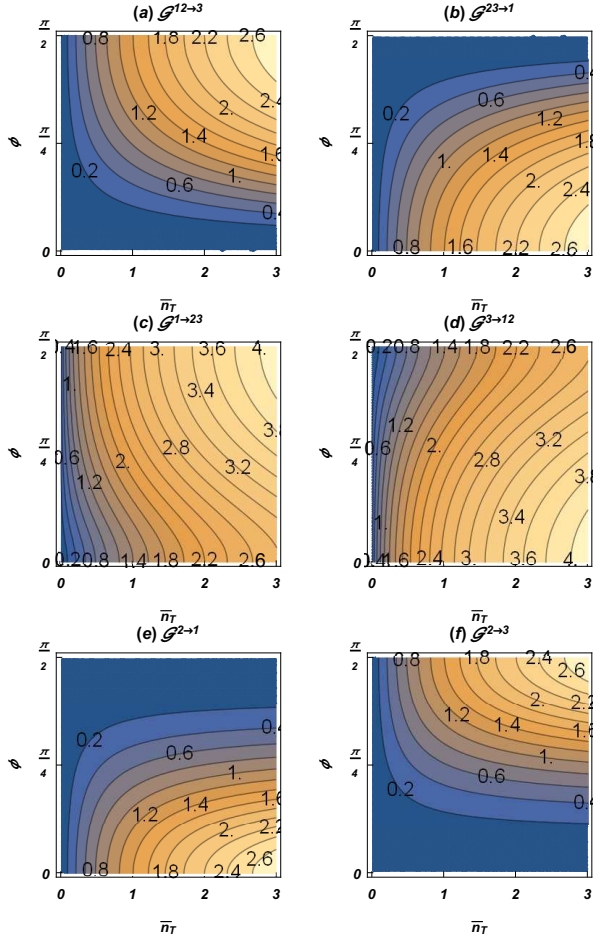}
\caption{(a) $\mathcal{G}^{12\rightarrow 3}$, (b) $\mathcal{G}%
^{23\rightarrow 1}$, (c) $\mathcal{G}^{1\rightarrow 23}$, (d) $\mathcal{G}%
^{3\rightarrow 12}$, (e) $\mathcal{G}^{2\rightarrow 1}$, and (f) $\mathcal{G}%
^{2\rightarrow 3}$ as functions of $\bar{n}_{T}$ and $\protect\phi $.}
\end{figure}
\begin{figure}[tbp]
\label{Fig4ab} {\centering\includegraphics[width=0.9\columnwidth]{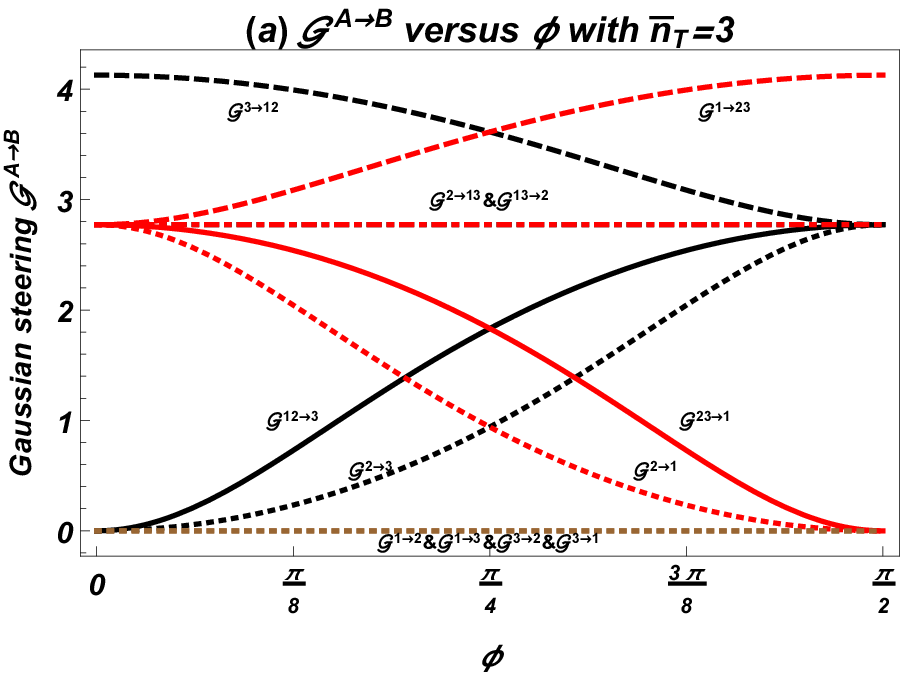}}
{\centering\includegraphics[width=0.9\columnwidth]{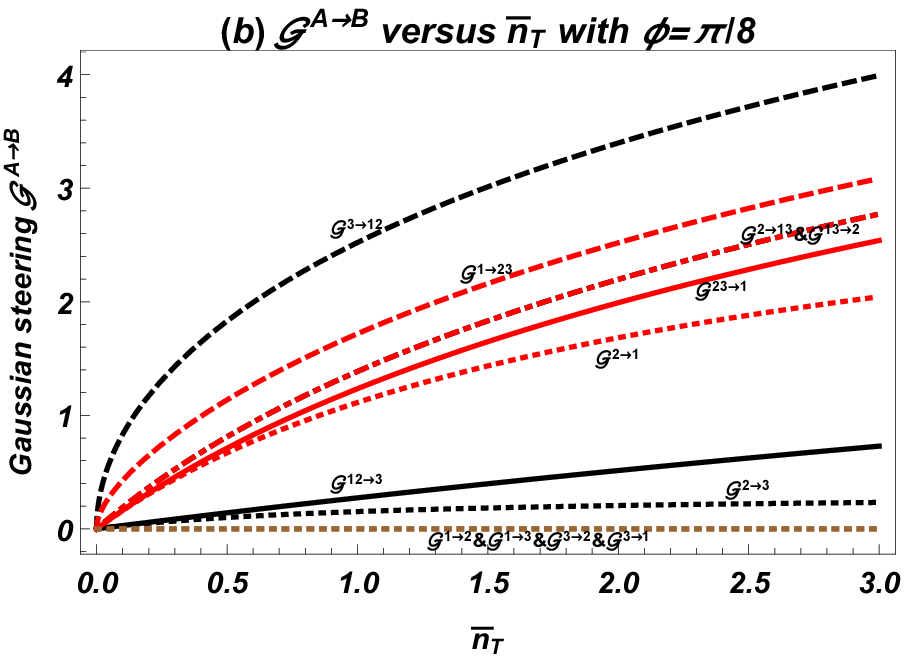}}
\caption{(a) $\mathcal{G}^{A\rightarrow B}$ versus $\protect\phi $ (setting $%
\bar{n}_{T}=3$). (b) $\mathcal{G}^{A\rightarrow B}$ versus $\bar{n}_{T}$
(setting $\protect\phi =\protect\pi /8$).}
\end{figure}

\subsection{Monogamy relations}

Monogamy means that two observers cannot simultaneously steer the state of
the third party. Both theoretical and experimental results show the
monogamous relation in multipartite EPR steering \cite{79}. In 2017, Xiang
\textit{et al.} defined the concept of the residual Gaussian steering (RGS)
\cite{77}. Here, we use the RGS to quantify the genuine tripartite steering
for the C3MSV. Using all the above expressions from Eqs.(\ref{2-5}) to (\ref%
{2-18}), we check that the following monogamy relations%
\begin{eqnarray}
\mathcal{G}^{\left( 23\right) \rightarrow 1}-\mathcal{G}^{2\rightarrow 1}-%
\mathcal{G}^{3\rightarrow 1} &\geq &0,  \notag \\
\mathcal{G}^{\left( 31\right) \rightarrow 2}-\mathcal{G}^{3\rightarrow 2}-%
\mathcal{G}^{1\rightarrow 2} &\geq &0,  \notag \\
\mathcal{G}^{\left( 12\right) \rightarrow 3}-\mathcal{G}^{1\rightarrow 3}-%
\mathcal{G}^{2\rightarrow 3} &\geq &0,  \label{2-19}
\end{eqnarray}%
and%
\begin{eqnarray}
\mathcal{G}^{1\rightarrow \left( 23\right) }-\mathcal{G}^{1\rightarrow 2}-%
\mathcal{G}^{1\rightarrow 3} &\geq &0,  \notag \\
\mathcal{G}^{2\rightarrow \left( 31\right) }-\mathcal{G}^{2\rightarrow 3}-%
\mathcal{G}^{2\rightarrow 1} &\geq &0,  \notag \\
\mathcal{G}^{3\rightarrow \left( 12\right) }-\mathcal{G}^{3\rightarrow 1}-%
\mathcal{G}^{3\rightarrow 2} &\geq &0,  \label{2-20}
\end{eqnarray}%
hold for the C3MSV. Further, we consider the RGS%
\begin{eqnarray}
\mathcal{G}^{1:2:3} &=&\min_{\left\langle i,j,k\right\rangle }\{\mathcal{G}%
^{\left( jk\right) \rightarrow i}-\mathcal{G}^{j\rightarrow i}-\mathcal{G}%
^{k\rightarrow i}\}  \notag \\
&=&\min_{\left\langle i,j,k\right\rangle }\{\mathcal{G}^{i\rightarrow \left(
jk\right) }-\mathcal{G}^{i\rightarrow j}-\mathcal{G}^{i\rightarrow k}\},
\label{2-21}
\end{eqnarray}%
for the C3MSV, where $\left\langle i,j,k\right\rangle $ denotes any cycle
permutation of $1$, $2$, and $3$. In Fig.5(a) we plot the RGS as a function
of $\bar{n}_{T}$\ and $\phi $. From which, we see that the RGS\ is maximized
on bisymmetric C3MSV with $\phi =\pi /4$, i.e., $r_{1}=r_{2}$. In this case,
the genuine tripartite $\mathcal{G}^{1:2:3}$\ reduces to the collective
steering $\mathcal{G}^{13\rightarrow 2}=\mathcal{G}^{2\rightarrow 13}=2\ln
(c^{2}+s^{2})$. Figure 5(b) presents the RGS as a function of $\phi $ with
different $\bar{n}_{T}$, which are the sections of Fig.5(a). Indeed, the RGS
acts as an indicator of collective steering-type correlations.
\begin{figure}[tbp]
\label{Fig5} \centering\includegraphics[width=0.9\columnwidth]{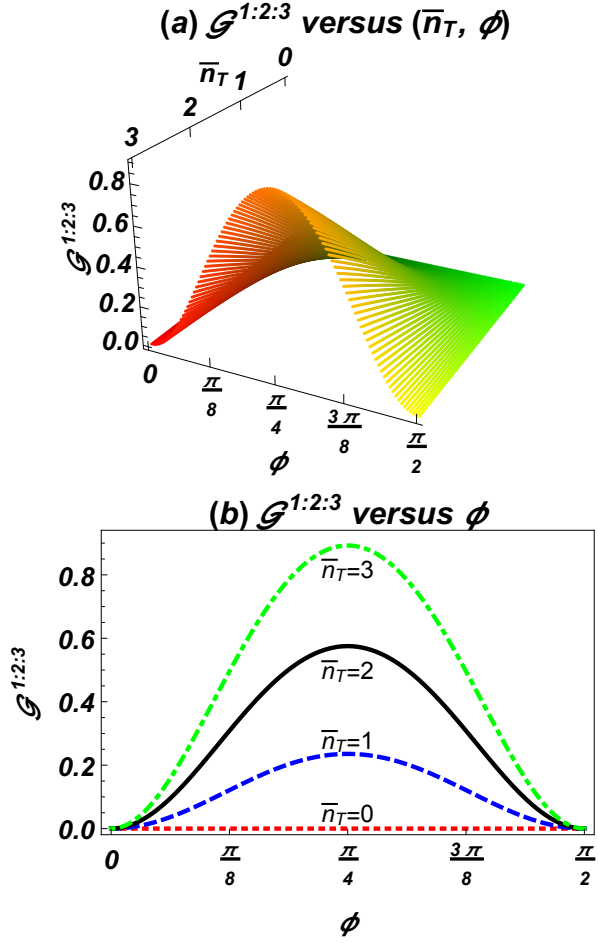}
\caption{RGS $\mathcal{G}^{1:2:3}$ for the C3MSV with CM $V$ (a) in the ($%
\bar{n}_{T}$, $\protect\phi $) space and (b) versus $\protect\phi $ with
different $\bar{n}_{T}$.}
\end{figure}

\section{Decoherence of steering for the C3MSV}

When dealing in a practical application, detector efficiencies and real
world effects such as losses and electronic noise will arise and become
crucial in a real experimental demonstration. Especially in the quantum
realm, decoherence properties will dominate.\ Following the handling ways of
Reid's group \cite{80} and Paris's group \cite{66,81,82}, we study the
decoherence of the steering for the C3MSV in this section. As shown in
Fig.6, we consider the evolution of the C3MSV\ in three independent noisy
channels (characterized by the loss rates $\gamma _{j}$\ and the thermal
photons $\bar{n}_{R_{j}}$). The solution in mode $j$ is straightforward to
evaluate by using the operator Langevin equation \cite{83,84}%
\begin{equation}
\dot{a}_{j}=-\gamma _{j}a_{j}+\sqrt{2\gamma _{j}}\Gamma _{j},  \label{3-1}
\end{equation}%
which describe the evolution of the mode operator $a_{j}$. Here, the
annihilation operator $\Gamma _{j}$\ describes the thermal reservoir $j$
with the occupation number $\bar{n}_{R_{j}}$ and the factor $\gamma _{j}$\
of mode $j$\ describes the decay (loss) rate that is induced by its
reservoir.
\begin{figure}[tbp]
\label{Fig6} \centering\includegraphics[width=0.8\columnwidth]{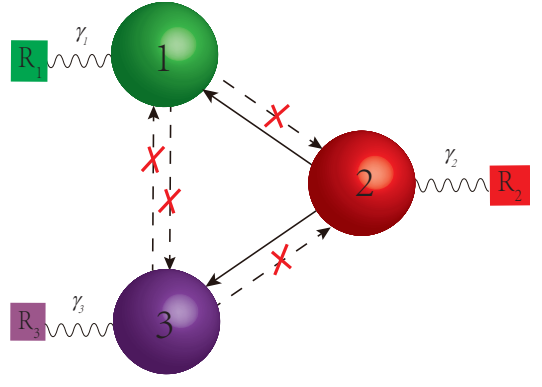}
\caption{Each mode (mode $j$) in the C3MSV is independently coupled to its
respective reservoir $R_{j}$ parametrized by the loss rate $\protect\gamma %
_{j}$. These three couplings induce the decoherence of the steerings of the
C3MSV.}
\end{figure}

Using the results provided in Appendix B, we can obtain the CM at time $t$
as follows
\begin{equation}
V\left( t\right) =\left(
\begin{array}{ccc}
\mathcal{A}_{1}\text{ }1_{2} & \mathcal{B}_{1}\text{ }\Sigma _{\theta _{1}}
& \mathcal{B}_{3}\text{ }R_{\theta _{2}-\theta _{1}} \\
\mathcal{B}_{1}\text{ }\Sigma _{\theta _{1}} & \mathcal{A}_{2}\text{ }1_{2}
& \mathcal{B}_{2}\text{ }\Sigma _{\theta _{2}} \\
\text{ }\mathcal{B}_{3}\text{ }\tilde{R}_{\theta _{2}-\theta _{1}} & \text{ }%
\mathcal{B}_{2}\Sigma _{\theta _{2}} & \mathcal{A}_{3}\text{ }1_{2}%
\end{array}%
\right) ,  \label{3-2}
\end{equation}%
with $\mathcal{A}_{j}=1+2\bar{n}_{j}+2\bar{n}_{R_{j}}(1-e^{-2\gamma _{j}t})$%
, $\mathcal{B}_{1}=-2sce^{-(\gamma _{1}+\gamma _{2})t}\cos \phi $, $\mathcal{%
B}_{2}=-2sce^{-(\gamma _{2}+\gamma _{3})t}\sin \phi $, and $\mathcal{B}%
_{3}=s^{2}e^{-(\gamma _{1}+\gamma _{3})t}\sin 2\phi $. Equation (\ref{3-2})
with $t=0$ can be reduced to Eq.(\ref{2-1}) as expected. Based on the CM in
Eq.(\ref{3-2}) and using the aforementioned steering criterion, we can
analyze the evolution of the steering.
\begin{figure*}[tbp]
\label{Fig7} \centering\includegraphics[width=1.8\columnwidth]{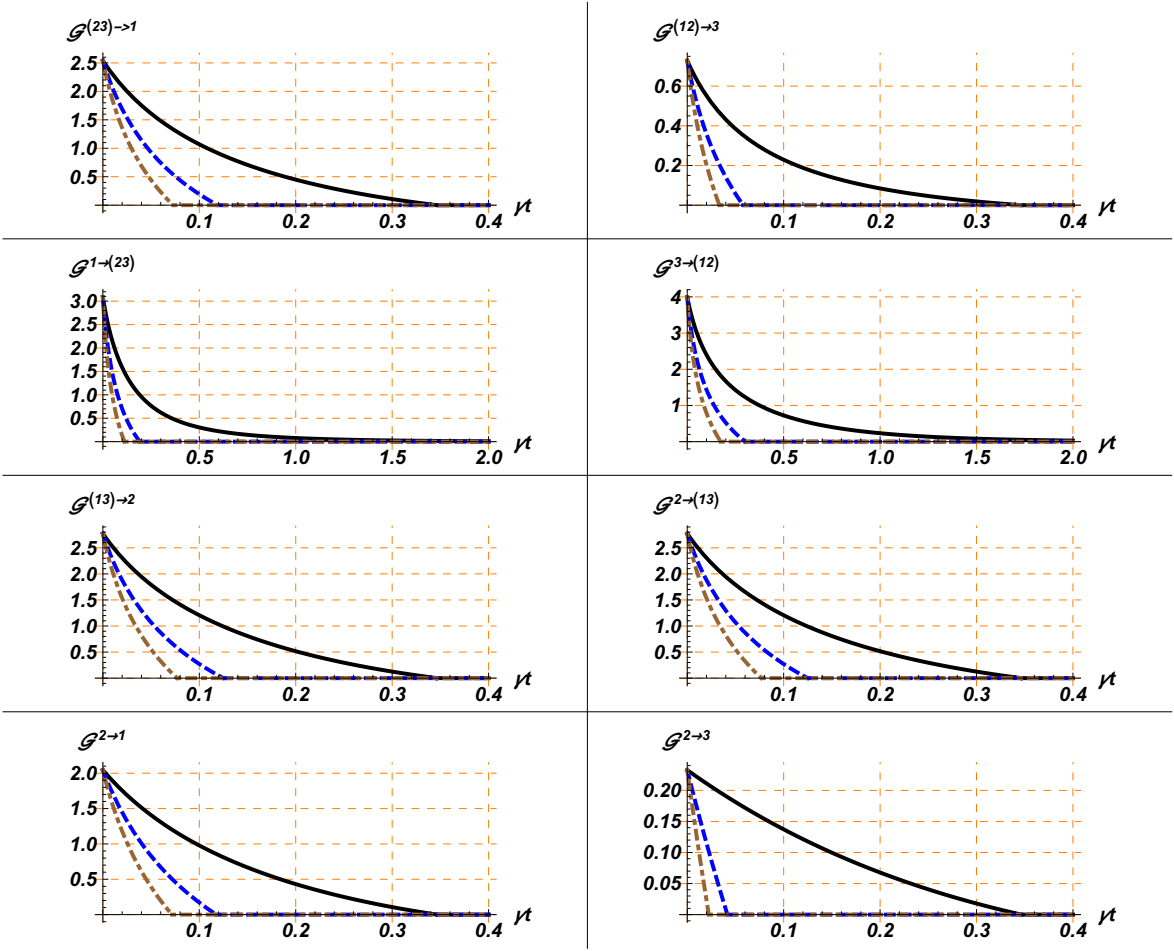}
\caption{Decoherence of several steerings of the C3MSV with $\bar{n}_{T}=3$
and $\protect\phi =\protect\pi /8$ and in different environments. $\mathcal{G%
}^{A\rightarrow B}$ versus $\protect\gamma t$\ in different $\bar{n}_{R}$.
The solid black line refers to the case $\bar{n}_{R}=0$. The dashed blue
line refers to the case $\bar{n}_{R}=0.5$. The dotdashed brown line refers
to the case $\bar{n}_{R}=1$. For each case, the sudden death will be
observed at a threshold time.}
\end{figure*}

Quite obviously, the dynamics of the steering is very complex because the
interaction is related with many parameters, including $r$, $\phi $, $\theta
_{1}$, $\theta _{2}$, $\gamma _{1}$, $\gamma _{2}$, $\gamma _{3}$, $\bar{n}%
_{R_{1}}$, $\bar{n}_{R_{2}}$, $\bar{n}_{R_{3}}$, and $t$. In fact, EPR
steering may be adjusted by varying the noise on different parties of the
C3MSV. Similar works on manipulating the direction \cite{85} or the dynamics
(such as death or revival) \cite{86} of EPR steering have been demonstrated.
Without loss of generality, we only set $\gamma _{1}=\gamma _{2}=\gamma
_{3}=\gamma $ and $\bar{n}_{R_{1}}=\bar{n}_{R_{2}}=\bar{n}_{R_{3}}=\bar{n}%
_{R}$. Using the C3MSV with $\bar{n}_{T}=3$ and $\phi =\pi /8$ and the
environments with $\bar{n}_{R}=0$, $0.5$, and $1$ as an example, we depict
the evolution of several steerings in Fig.7. These results show that: (i)
The steerability will decrease as time $\gamma t$ increases, and (ii) Until $%
\gamma t$ exceeds a certain threshold value, sudden death is observed.
Moreover, the threshold time is shorten by increasing $\bar{n}_{R}$. Taking $%
\mathcal{G}^{23\rightarrow 1}$ of Fig.7 as an example, the sudden deaths are
observed at $\gamma t$ $=$ $0.346574$, $0.11903$, and $0.0729227$, for $\bar{%
n}_{R}=0$, $0.5$, and $1$, respectively.

\section{\textbf{Protocols of preparing Wigner negativity} remotely}

As Walschaers \textit{et al.} recently pointed out, when party $A$ and party
$B$ share a Gaussian state, party $B$ can perform some measurement on itself
to create Wigner negativity on party $A$, if and only if there is a Gaussian
steering from party $A$ to party $B$ \cite{57}. Moreover, they provided an
intuitive method to quantify remotely generated WN by employing non-Gaussian
operation of photon subtraction. Following methods in Walschaers' work \cite%
{59} and Xiang's work \cite{62}, we investigate the remote creation and
distribution of WN in the tripartite C3MSV. Here, we declare that we only
study ideal and conceptual protocols of preparing WN, without considering
any lossy channels. Based on the C3MSV, we keep the steered party $B$ in the
local station and send the steering party $A$ to the remote position. After
appropriate single-photon subtraction(s) on the steered party $B$, the
steering party $A$ becomes a reduced non-Gaussian state $\rho _{B_{a}|A}$.
In some cases, we can generate Wigner negative states in the remote
position. For state $\rho _{j}$, we can derive its Wigner function (WF) by $%
W_{\rho _{j}}\left( \beta _{j}\right) =\mathrm{Tr}(\hat{O}_{w_{j}}\rho _{j})$%
, with $\hat{O}_{w_{j}}=\frac{2}{\pi }:e^{-2(\hat{a}_{j}^{\dag }-\beta
_{j}^{\ast })(\hat{a}_{j}-\beta _{j})}:$ ($:\cdots :$ denotes the normal
ordering) and $\beta _{j}=\left( x_{j}+iy_{j}\right) /\sqrt{2}$\cite{87,88}.
Furthermore, we can quantify the WN of $\rho _{B_{a}|A}$ as
\begin{equation}
\mathcal{N}\equiv \int \left\vert W(\beta )\right\vert d^{2n_{A}}\beta -1,
\label{4-1}
\end{equation}%
with $\beta \in
%TCIMACRO{\U{211d} }%
%BeginExpansion
\mathbb{R}
%EndExpansion
^{2n_{A}}$, where $n_{A}$\ is the mode number considered in party $A$. As
shown schematically in Fig.8, we propose protocols of generating 18 kinds of
$\rho _{B_{a}|A}$s, whose analytical WFs are given in Appendix \textbf{C}.
As examples, we plot WFs for $\rho _{B_{a}|A}$s with $\bar{n}_{T}=3$ and $%
\phi =\pi /8$ in Fig.9, where only several WFs exhibits the WNs.\
\begin{figure*}[tbp]
\label{Fig8} \centering\includegraphics[width=1.8\columnwidth]{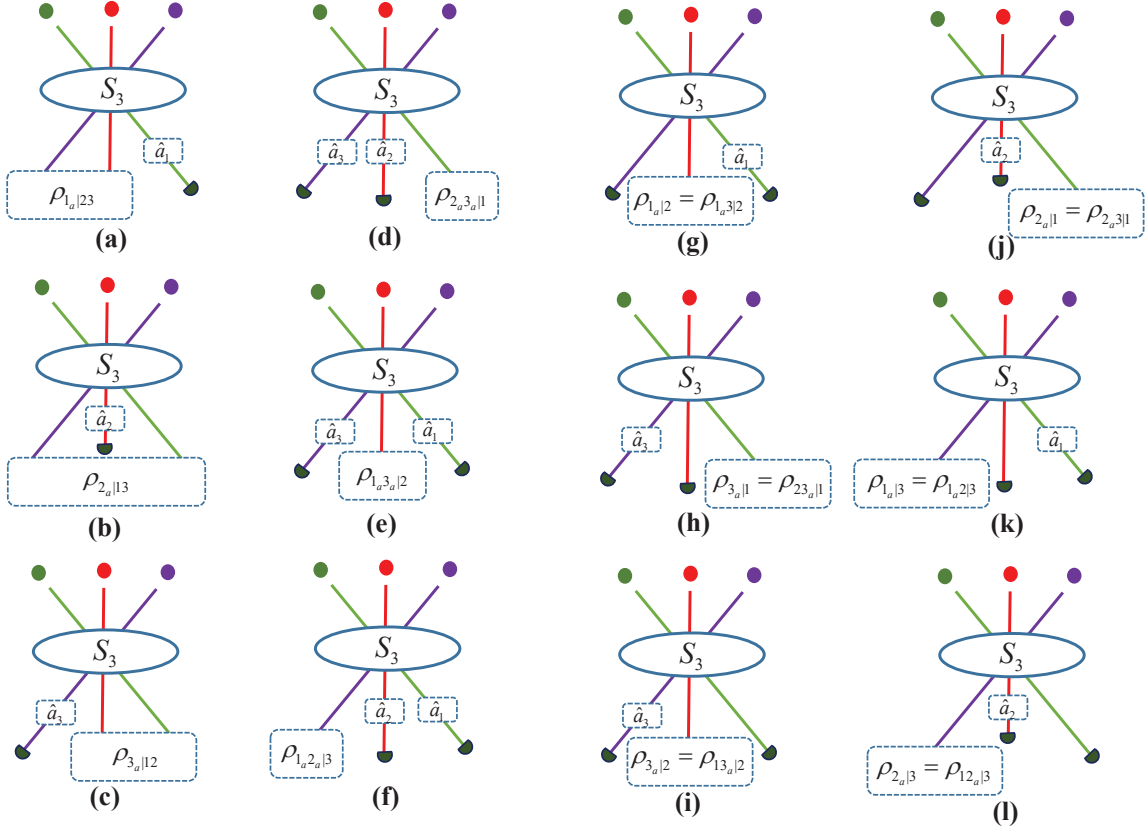}
\caption{Ideal and conceptual schemes of remote generated non-Gaussian
states $\protect\rho _{B_{a}|A}$ based on the C3MSV, without considering the
loss in any channel. Note that cases (a)-(c) are two-mode states and cases
(d)-(l) are one-mode states.}
\end{figure*}
\begin{figure*}[tbp]
\label{Fig9} \centering\includegraphics[width=1.8\columnwidth]{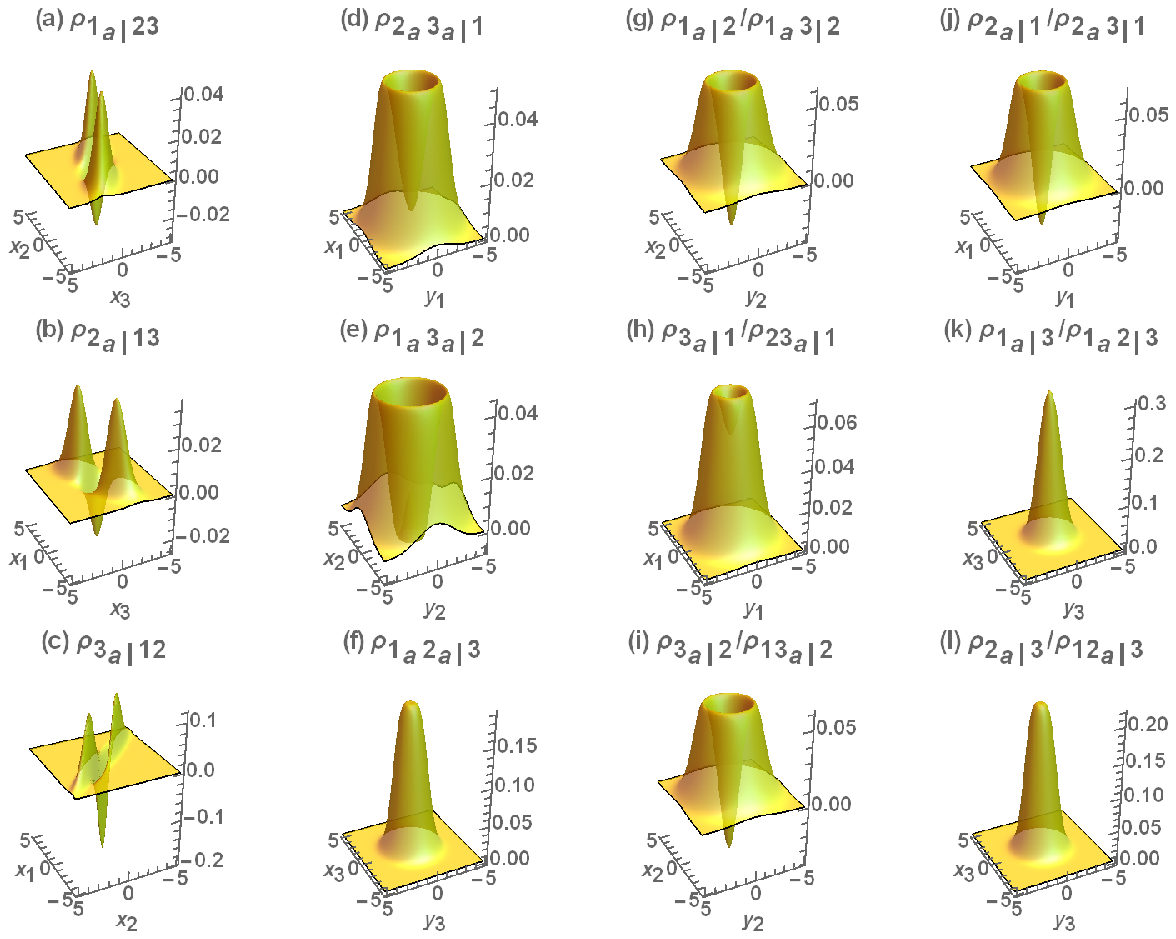}
\caption{WFs of $\protect\rho _{B_{a}|A}$s corresponding to Fig.8, with $%
\bar{n}_{T}=3$, $\protect\phi =\protect\pi /8$, and $\protect\theta _{1}=%
\protect\theta _{2}=0$. Some cases have WNs and some cases have no WNs.}
\end{figure*}

Indeed, the amount of WN cannot be freely distributed among different modes.
It can be influenced by the considered protocols and the interaction
parameters. In order to explain the characters, we plot some WNs $\mathcal{N}
$\ versus $\phi $ by fixing $\bar{n}_{T}=3$ in Fig.10 and\ versus $\bar{n}%
_{T}$ by fixing $\phi =\pi /8$ in Fig.11. The details are explained as
follows.
\begin{figure*}[tbp]
\label{Fig10} \centering\includegraphics[width=1.8\columnwidth]{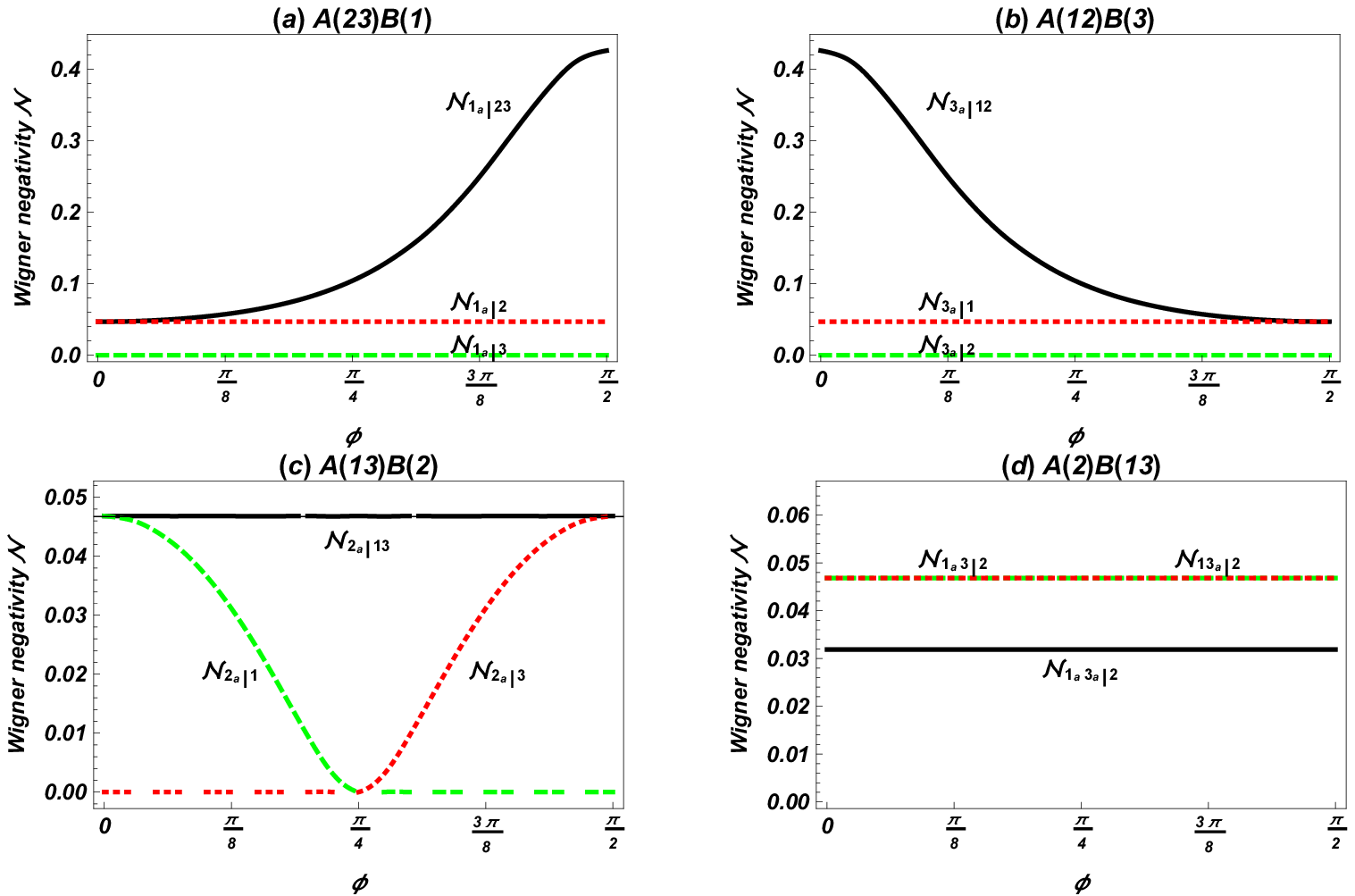}
\caption{WNs versus $\protect\phi $ with fixed $\bar{n}_{T}=3$, for some $%
\protect\rho _{B_{a}|A}$s in cases (a) $A(23)B(1)$, (b) $A(12)B(3)$, (c) $%
A(13)B(2)$, and (d) $A(2)B(13)$.}
\end{figure*}
\begin{figure*}[tbp]
\label{Fig11} \centering\includegraphics[width=1.8\columnwidth]{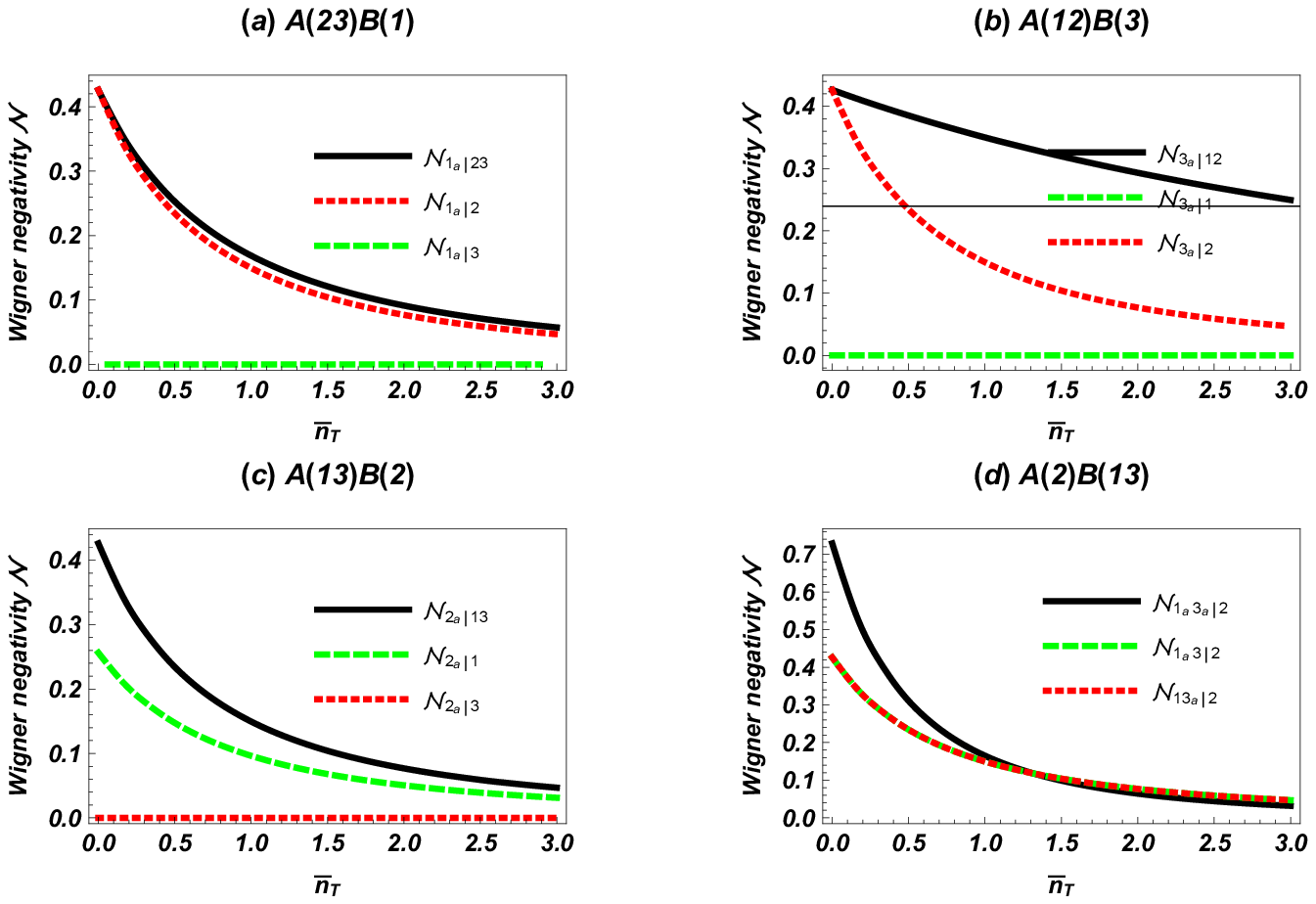}
\caption{WNs versus $\bar{n}_{T}$ with fixed $\protect\phi =\protect\pi /8$,
for some $\protect\rho _{B_{a}|A}$s in cases (a) $A(23)B(1)$, (b) $A(12)B(3)$%
, (c) $A(13)B(2)$, and (d) $A(2)B(13)$.}
\end{figure*}

\textit{Case }$A(23)B(1)$\textit{:} In this case, the steering party $A$
includes mode 2 and mode 3 and the steered party $B$ includes mode 1.
Performing appropriate photon subtraction(s) in the local position, we can
remotely generate the following states with their respective WNs:
\begin{eqnarray}
\rho _{1_{a}|23} &=&\mathrm{Tr}_{1}\left( \rho _{1_{a}23}\right) \rightarrow
\mathcal{N}_{1_{a}|23},  \notag \\
\rho _{1_{a}|2} &=&\mathrm{Tr}_{1,3}\left( \rho _{1_{a}23}\right)
\rightarrow \mathcal{N}_{1_{a}|2},  \notag \\
\rho _{1_{a}|3} &=&\mathrm{Tr}_{1,2}\left( \rho _{1_{a}23}\right)
\rightarrow \mathcal{N}_{1_{a}|3},  \label{4-2}
\end{eqnarray}%
where $\rho _{1_{a}23}=\left\vert \epsilon _{1}\right\vert _{1}^{-2}\hat{a}%
_{1}\rho _{123}\hat{a}_{1}^{\dag }$.

We plot $\mathcal{N}_{1_{a}|23}$, $\mathcal{N}_{1_{a}|2}$, and $\mathcal{N}%
_{1_{a}|3}$ as functions of $\phi $ in Fig.10(a) and as functions of $\bar{n}%
_{T}$ in Fig.11(a). From these figures, we find that WNs are generated
remotely in the group (23), mode 2 and mode 3, respectively, after a
single-photon subtraction on mode 1. Moreover, we see that (i) $\mathcal{N}%
_{1_{a}|23}$ is a monotonically increasing function of $\phi $ from $0.04682$%
\ at $\phi =0$\ to $0.42614$\ at $\phi =\pi /2$; (ii) $\mathcal{N}%
_{1_{a}|2}\ $ remains as $0.04682$ for any $\phi $; (iii) $\mathcal{N}%
_{1_{a}|3}\ $ remains as $0$ for any $\phi $; (iv) $\mathcal{N}%
_{1_{a}|23}\geq \mathcal{N}_{1_{a}|2}+\mathcal{N}_{1_{a}|3}$; and (v) as $%
\bar{n}_{T}$ is increasing, all these WNs will be limited to $0$.

\textit{Case }$A(12)B(3)$\textit{:} In this case, the steering party $A$
include mode 1 and mode 2 and the steered party $B$ include mode 3.
Performing appropriate photon subtraction(s) in the local position, we can
remotely generate the following states with their respective WNs:
\begin{eqnarray}
\rho _{3_{a}|12} &=&\mathrm{Tr}_{3}\left( \rho _{123_{a}}\right) \rightarrow
\mathcal{N}_{3_{a}|12},  \notag \\
\rho _{3_{a}|1} &=&\mathrm{Tr}_{2,3}\left( \rho _{123_{a}}\right)
\rightarrow \mathcal{N}_{3_{a}|1},  \notag \\
\rho _{3_{a}|2} &=&\mathrm{Tr}_{1,3}\left( \rho _{123_{a}}\right)
\rightarrow \mathcal{N}_{3_{a}|2},  \label{4-3}
\end{eqnarray}%
where $\rho _{123_{a}}=\left\vert \epsilon _{2}\right\vert ^{-2}\hat{a}%
_{3}\rho _{123}\hat{a}_{3}^{\dag }$.

We plot $\mathcal{N}_{3_{a}|12}$, $\mathcal{N}_{3_{a}|1}$, and $\mathcal{N}%
_{3_{a}|2}$ as functions of $\phi $ in Fig.10(b) and as functions of $\bar{n}%
_{T}$ in Fig.11(b). From these figures, we find that WNs are generated
remotely in the group (12), mode 1 and mode 2, respectively, after a
single-photon subtraction on mode 3. Moreover, we see that (i) $\mathcal{N}%
_{3_{a}|12}$ is a monotonically decreasing function of $\phi $ from $0.42614$%
\ at $\phi =0$\ to $0.04682$\ at $\phi =\pi /2$; (ii) $\mathcal{N}%
_{3_{a}|2}\ $ remains as $0.04682$ for any $\phi $; (iii) $\mathcal{N}%
_{3_{a}|1}\ $ remains as $0$ for any $\phi $; (iv) $\mathcal{N}%
_{3_{a}|12}\geq \mathcal{N}_{3_{a}|1}+\mathcal{N}_{3_{a}|2}$; and (v) as $%
\bar{n}_{T}$ is increasing, all these WNs will be limited to $0$.

\textit{Case }$A(13)B(2)$\textit{:} In this case, the steering party $A$
includes mode 1 and mode 3 and the steered party $B$ includes mode 2.
Performing appropriate photon subtraction(s) in the local position, we can
remotely generate the following states with their respective WNs:
\begin{eqnarray}
\rho _{2_{a}|13} &=&\mathrm{Tr}_{2}\left( \rho _{12_{a}3}\right) \rightarrow
\mathcal{N}_{2_{a}|13},  \notag \\
\rho _{2_{a}|1} &=&\mathrm{Tr}_{2,3}\left( \rho _{12_{a}3}\right)
\rightarrow \mathcal{N}_{2_{a}|1},  \notag \\
\rho _{2_{a}|3} &=&\mathrm{Tr}_{1,2}\left( \rho _{12_{a}3}\right)
\rightarrow \mathcal{N}_{2_{a}|3},  \label{4-4}
\end{eqnarray}%
where $\rho _{12_{a}3}=s^{-2}\hat{a}_{2}\rho _{123}\hat{a}_{2}^{\dag }$.

We plot $\mathcal{N}_{2_{a}|13}$, $\mathcal{N}_{2_{a}|1}$, and $\mathcal{N}%
_{2_{a}|3}$ as functions of $\phi $ in Fig.10(c) and as functions of $\bar{n}%
_{T}$ in Fig.11(c). From these figures, we find that WNs are generated
remotely in the group (13), mode 1 and mode 3, respectively, after a
single-photon subtraction on mode 2. Moreover, we see that (i) $\mathcal{N}%
_{2_{a}|13}$ remains as $0.4683$ for any $\phi $; (ii) $\mathcal{N}%
_{2_{a}|1} $ decreases from $0.4683$\ to $0$ in $\left[ 0,\pi /4\right] $
and remains as $0$ in $\left[ \pi /4,\pi /2\right] $; (iii) $\mathcal{N}%
_{2_{a}|3}$ remains as $0$ in $\left[ 0,\pi /4\right] $ and increases from $%
0 $\ to $0.4683$ in $\left[ \pi /4,\pi /2\right] $; (iv) $\mathcal{N}%
_{2_{a}|13}\geq \mathcal{N}_{2_{a}|1}+\mathcal{N}_{2_{a}|3}$; and (v) as $%
\bar{n}_{T}$ is increasing, all these WNs will be limited to $0$.

\textit{Case }$A(2)B(13)$\textit{:} In this case, the steering party $A$
include mode 2 and the steered party $B$ include mode 1 and mode 3.
Performing appropriate photon subtraction(s) in the local position, we can
remotely generate the following states with their respective WNs:
\begin{eqnarray}
\rho _{1_{a}3_{a}|2} &=&\mathrm{Tr}_{1,3}\left( \rho _{1_{a}23_{a}}\right)
\rightarrow \mathcal{N}_{1_{a}3_{a}|2},  \notag \\
\rho _{1_{a}3|2} &=&\mathrm{Tr}_{1,3}\left( \rho _{1_{a}23}\right)
\rightarrow \mathcal{N}_{1_{a}3|2},  \notag \\
\rho _{13_{a}|2} &=&\mathrm{Tr}_{1,3}\left( \rho _{123_{a}}\right)
\rightarrow \mathcal{N}_{13_{a}|2},  \label{4-5}
\end{eqnarray}%
where $\rho _{1_{a}23_{a}}=\frac{1}{2}\left\vert \epsilon _{1}\epsilon
_{2}\right\vert ^{-2}\hat{a}_{1}\hat{a}_{3}\rho _{123}\hat{a}_{1}^{\dag }%
\hat{a}_{3}^{\dag }$.

We plot $\mathcal{N}_{1_{a}3_{a}|2}$, $\mathcal{N}_{1_{a}3|2}$, and $%
\mathcal{N}_{13_{a}|2}$ as functions of $\phi $ in Fig.10(d) and as
functions of $\bar{n}_{T}$ in Fig.11(d). From these figures, we find that
WNs are generated remotely in mode 2, after single-photon subtractions on
each mode of the group (13) simultaneously, or after a single-photon
subtraction on mode 1 or mode 3, respectively. Here, we see that (1) $%
\mathcal{N}_{1_{a}3_{a}|2}$ remains as $0.0318528$, (2) $\mathcal{N}%
_{1_{a}3|2}$ and $\mathcal{N}_{13_{a}|2}\ $remain as $0.04683$; (3) $%
\mathcal{N}_{1_{a}3_{a}|2}<\mathcal{N}_{1_{a}3|2}+\mathcal{N}_{13_{a}|2}$;
(4) As $\bar{n}_{T}$ increasing, all these WNs will limit to $0$. However,
although $\mathcal{G}^{2\rightarrow 1}>0$, $\mathcal{G}^{2\rightarrow 3}>0$,
and $\mathcal{G}^{2\rightarrow \left( 31\right) }>0$, we cannot achieve more
significant increase of the WNs in mode 2, after performing a single-photon
subtraction on each of mode 1 and mode 3.

\textit{Case }$A(1)B(23)$\textit{:} In this case, the steering party $A$
include mode 1 and the steered party $B$ include mode 2 and mode 3.
Performing appropriate photon subtraction(s) in the local position, we can
remotely generate the following states with their respective WNs:
\begin{eqnarray}
\rho _{2_{a}3_{a}|1} &=&\mathrm{Tr}_{2,3}\left( \rho _{12_{a}3_{a}}\right)
\rightarrow \mathcal{N}_{2_{a}3_{a}|1}\equiv 0,  \notag \\
\rho _{2_{a}3|1} &=&\mathrm{Tr}_{2,3}\left( \rho _{12_{a}3}\right)
\rightarrow \mathcal{N}_{2_{a}3|1}\equiv 0,  \notag \\
\rho _{23_{a}|1} &=&\mathrm{Tr}_{2,3}\left( \rho _{123_{a}}\right)
\rightarrow \mathcal{N}_{23_{a}|1}\equiv 0,  \label{4-6}
\end{eqnarray}%
where $\rho _{12_{a}3_{a}}=\left( c^{2}+s^{2}\right) ^{-1}\left\vert
\epsilon _{2}\right\vert ^{-2}\hat{a}_{2}\hat{a}_{3}\rho _{123}\hat{a}%
_{2}^{\dag }\hat{a}_{3}^{\dag }$. For any $\bar{n}_{T}$\ and $\phi $, we see
$\mathcal{N}_{2_{a}3_{a}|1}=\mathcal{N}_{2_{a}3|1}=\mathcal{N}%
_{23_{a}|1}\equiv 0$. That is to say, no WNs are generated remotely in mode
1, after single-photon subtractions on each mode of the group (23)
simultaneously, or after a single-photon subtraction on mode 2 or mode 3.
Surprisingly, $\mathcal{N}_{2_{a}3_{a}|1}=0$ although $\mathcal{G}%
^{1\rightarrow 23}>0$.

\textit{Case }$A(3)B(12)$\textit{:} In this case, the steering party $A$
include mode-3 and the steered party $B$ include mode 1 and mode 2.
Performing appropriate photon subtraction(s) in the local position, we can
remotely generate the following states with their respective WNs:
\begin{eqnarray}
\rho _{1_{a}2_{a}|3} &=&\mathrm{Tr}_{1,2}\left( \rho _{1_{a}2_{a}3}\right)
\rightarrow \mathcal{N}_{1_{a}2_{a}|3}\equiv 0,  \notag \\
\rho _{1_{a}2|3} &=&\mathrm{Tr}_{1,2}\left( \rho _{1_{a}23}\right)
\rightarrow \mathcal{N}_{1_{a}2|3}\equiv 0,  \notag \\
\rho _{12_{a}|3} &=&\mathrm{Tr}_{1,2}\left( \rho _{12_{a}3}\right)
\rightarrow \mathcal{N}_{12_{a}|3}\equiv 0,  \label{4-7}
\end{eqnarray}%
where $\rho _{1_{a}2_{a}3}=\left( c^{2}+s^{2}\right) ^{-1}\left\vert
\epsilon _{1}\right\vert ^{-2}\hat{a}_{1}\hat{a}_{2}\rho _{123}\hat{a}%
_{1}^{\dag }\hat{a}_{2}^{\dag }$. For any $\bar{n}_{T}$\ and $\phi $, we see
$\mathcal{N}_{1_{a}2_{a}|3}=\mathcal{N}_{1_{a}2|3}=\mathcal{N}%
_{12_{a}|3}\equiv 0$. That is to say, no WNs are generated remotely in mode
3, after single-photon subtractions on each mode of the group (12)
simultaneously, or after a single-photon subtraction on mode 1 or mode 2.
Surprisingly, $\mathcal{N}_{1_{a}2_{a}|3}=0$ although $\mathcal{G}%
^{3\rightarrow 12}>0$.

So far, we have quantified all remotely generated WNs in terms of Eq.(\ref%
{4-1}). It is obvious to see that the amount of WN cannot be freely
distributed among different modes.

\section{Conclusion and discussion}

To summarize, we studied the C3MSV and showed how it can be used for
steering. By taking different bipartite assignment in the C3MSV, we
investigated all bipartite\ Gaussian steerings present in the C3MSV. These
steerings include no steering, one-way steering and two-way steering.
Moreover, the steerability can be adjusted by the interaction parameters. In
addition, we also studied the decoherence of the steering for the C3MSV and
found that the steering will die suddenly at a threshold time. Using the
C3MSV as the resource, we proposed conceptual schemes to remotely generate
Wigner negative states. We analyzed and compared the distributions of the
Gaussian steering and the WNs over different modes. Normally, one expect
that stronger steerability induces more WN. That is, if $\mathcal{G}%
^{A\rightarrow B}>0$, then $\mathcal{N}>0$; and if $\mathcal{G}%
^{A\rightarrow B}=0$, then $\mathcal{N}=0$.\ But this is not the case for
the C3MSV. For example, although $\mathcal{G}^{1\rightarrow 23}>0$\ and $%
\mathcal{G}^{3\rightarrow 12}>0$, $\rho _{2_{a}3_{a}|1}$\ and $\rho
_{1_{a}2_{a}|3}$\ cannot exhibit WN. These results further verify that
quantum correlations are not always a necessary requirement for the
conditional generation of WN\cite{57}.

People expect that the correlations can be more robust to environmental
influences (including loss and noise) \cite{89,90}. Meanwhile, measurement
will have nonunity detection efficiency \cite{91,92} accompanied with
information leakage \cite{93}. With the help of squeezed states \cite{94}
and erasure corrections \cite{95}, one can establish quantum optical
coherence over longer distances to diminish the effect from losses and
noises. In the aspects of\ experiment and measurement, our paper is a
reservoir of more discussions. Although our work is theoretical and ideal,
we still believe that our results may also lay a solid theoretical
foundation for a future practical study.

Practical quantum communications (including quantum internet \cite{96,97},
satellite communication \cite{98}, and online banking \cite{99}) require
multipartite correlation and high security \cite{100,101}. Fortunately, all
these problems will solved by using protocols involved in quantum steering
\cite{102}.\ Specific properties of the C3MSV (including squeezing,
entanglement and steering) have laid a good foundation for applications in
quantum technologies. So, we believe the C3MSV will become a useful
entangled resource in future quantum communication. For example, following
previous works\cite{65,66,103} and using the C3MSV, one can construct a new
scheme to\ teleclone pure Gaussian states.

\begin{acknowledgments}
This paper was supported by the National Natural Science Foundation of China
(Grant No. 11665013).
\end{acknowledgments}

\section*{\textbf{Appendix A: The C3MSO and the C3MSV}}

In this appendix, we give the transformation relation and the normal
ordering form for the C3MSO. In addition, we give the general expression to
calculate the expectation values we want for the C3MSV.

\textit{About the C3MSO. }Using the formula of Bogoliubov transformation, we
obtain the following transformation relations%
\begin{equation}
S_{3}A^{\dag }S_{3}^{\dag }=A^{\dag }P^{\ast }+AL^{\ast },S_{3}AS_{3}^{\dag
}=AP+A^{\dag }L,  \tag{A.1}
\end{equation}%
where $A^{\dag }=\left( \hat{a}_{1}^{\dag },\hat{a}_{2}^{\dag },\hat{a}%
_{3}^{\dag }\right) $, $A=\left( \hat{a}_{1},\hat{a}_{2},\hat{a}_{3}\right) $%
, and%
\begin{equation}
P=\left(
\begin{array}{ccc}
\kappa _{1} & 0 & \tau \\
0 & c & 0 \\
\tau ^{\ast } & 0 & \kappa _{2}%
\end{array}%
\right) ,L=\left(
\begin{array}{ccc}
0 & \epsilon _{1} & 0 \\
\epsilon _{1} & 0 & \epsilon _{2} \\
0 & \epsilon _{2} & 0%
\end{array}%
\right) .  \tag{A.2}
\end{equation}%
Here we set $\kappa _{1}=\sin ^{2}\phi +c\cos ^{2}\phi $, $\kappa _{2}=\cos
^{2}\phi +c\sin ^{2}\phi $, and $\tau =\frac{1}{2}(c-1)e^{i(\theta
_{2}-\theta _{1})}\sin 2\phi $.

According to the rule provided by Fan and co-workers\cite{104,105,106}, we
immediately obtain the normal ordering form of $S_{3}$ as follows%
\begin{align}
S_{3}& =\frac{1}{\sqrt{\det P}}e^{-\frac{1}{2}A^{\dag }(LP^{-1})\tilde{A}%
^{\dag }}  \notag \\
& :e^{A^{\dag }(\tilde{P}^{-1}-I)\tilde{A}}:e^{\frac{1}{2}A(P^{-1}L^{\ast })%
\tilde{A}}.  \tag{A.3}
\end{align}%
Of course, we can further use $:e^{A^{\dag }(\tilde{P}^{-1}-I)\tilde{A}%
}:=e^{A^{\dag }(\ln \tilde{P}^{-1})\tilde{A}}$ in above expression, where $%
\tilde{P}$\ denotes the transpose of $P$.

\textit{About the C3MSV. }Here, we give the following general expression of
expectation value:

\begin{align}
& \left\langle \hat{a}_{1}^{\dag k_{1}}\hat{a}_{2}^{\dag k_{2}}\hat{a}%
_{3}^{\dag k_{3}}\hat{a}_{1}^{l_{1}}\hat{a}_{2}^{l_{2}}\hat{a}%
_{3}^{l_{3}}\right\rangle  \notag \\
& =\partial _{\mu _{1}}^{k_{1}}\partial _{\mu _{2}}^{k_{2}}\partial _{\mu
_{3}}^{k_{3}}\partial _{\nu _{1}}^{l_{1}}\partial _{\nu
_{2}}^{l_{2}}\partial _{\nu _{3}}^{l_{3}}  \notag \\
& e^{\left\vert \epsilon _{1}\right\vert ^{2}\mu _{1}\nu _{1}+s^{2}\mu
_{2}\nu _{2}+\left\vert \epsilon _{2}\right\vert ^{2}\mu _{3}\nu
_{3}+\epsilon _{1}^{\ast }\epsilon _{2}\mu _{1}\nu _{3}+\epsilon
_{1}\epsilon _{2}^{\ast }\mu _{3}\nu _{1}}  \notag \\
& e^{-c\epsilon _{1}^{\ast }\mu _{1}\mu _{2}-c\epsilon _{1}\nu _{1}\nu
_{2}-c\epsilon _{2}^{\ast }\mu _{2}\mu _{3}-c\epsilon _{2}\nu _{2}\nu _{3}}
\notag \\
& |_{\mu _{1}=\mu _{2}=\mu _{3}=\nu _{1}=\nu _{2}=\nu _{3}=0},  \tag{A.4}
\end{align}%
from which one can study the statistical properties for the C3MSV. Notice
that $k_{1}$, $k_{2}$, $k_{3}$, $l_{1}$, $l_{2}$, and $l_{3}$\ are
non-negative integers.

\section*{\textbf{Appendix B: Derivation of evolution relation in the
reservoir}}

Using the Laplace transformation%
\begin{equation}
\tilde{a}_{j}\left( p\right) =\mathrm{LT}\left[ a_{j}\left( t\right) \right]
=\int_{0}^{\infty }dt\exp \left( -pt\right) a_{j}\left( t\right) ,  \tag{B1}
\end{equation}%
and $\mathrm{LT}\left[ \dot{a}_{j}\left( t\right) \right] =p\tilde{a}%
_{j}\left( p\right) -a_{j}\left( 0\right) $, Eq.(\ref{3-1}) yields%
\begin{equation}
\tilde{a}_{j}\left( p\right) =\frac{1}{p+\gamma _{j}}a_{j}\left( 0\right) +%
\sqrt{2\gamma _{j}}\frac{\tilde{\Gamma}_{j}\left( p\right) }{p+\gamma _{j}}
\tag{B2}
\end{equation}%
which leads to \cite{80}%
\begin{equation}
a_{j}\left( t\right) =e^{-\gamma _{j}t}a_{j}\left( 0\right) +\sqrt{2\gamma
_{j}}\int_{0}^{t}e^{-\gamma _{j}(t-\tau )}\Gamma _{j}\left( \tau \right)
d\tau .  \tag{B3}
\end{equation}

Notice that the quantum reservoir operators\ have correlations given by $%
\left\langle \Gamma _{j}\right\rangle =\left\langle \Gamma _{j}^{\dag
}\right\rangle =\left\langle \Gamma _{j}^{2}\right\rangle =\left\langle
\Gamma _{j}^{\dag 2}\right\rangle =0$ and%
\begin{align}
\left\langle \Gamma _{j}^{\dag }\left( \tau ^{\prime }\right) \Gamma
_{j}\left( \tau \right) \right\rangle & =\bar{n}_{R_{j}}\delta \left( \tau
^{\prime }-\tau \right)  \notag \\
\left\langle \Gamma _{j}\left( \tau \right) \Gamma _{j}^{\dag }\left( \tau
^{\prime }\right) \right\rangle & =(\bar{n}_{R_{j}}+1)\delta \left( \tau
^{\prime }-\tau \right) .  \tag{B4}
\end{align}%
as well as $\left\langle \Gamma _{j}\Gamma _{k}\right\rangle =\left\langle
\Gamma _{j}^{\dag }\Gamma _{k}^{\dag }\right\rangle =\left\langle \Gamma
_{j}\Gamma _{k}^{\dag }\right\rangle =\left\langle \Gamma _{j}^{\dag }\Gamma
_{k}\right\rangle =0$ for $j\neq k$. Thus we can calculate the moments at a
later time in terms of the initial moments, in terms of the relations such
as $\left\langle a_{j}\left( t\right) \right\rangle =e^{-\gamma
_{j}t}\left\langle a_{j}\left( 0\right) \right\rangle $, $\left\langle
a_{j}^{2}\left( t\right) \right\rangle =e^{-2\gamma _{j}t}\left\langle
a_{j}^{2}\left( 0\right) \right\rangle $, and%
\begin{align}
\left\langle a_{j}^{\dag }\left( t\right) a_{j}\left( t\right) \right\rangle
& =e^{-2\gamma _{j}t}\left\langle a_{j}^{\dag }\left( 0\right) a_{j}\left(
0\right) \right\rangle  \notag \\
& +\bar{n}_{R_{j}}\left( 1-e^{-2\gamma _{j}t}\right)  \tag{B5}
\end{align}%
for the same mode, as well as%
\begin{align}
\left\langle a_{j}\left( t\right) a_{k}\left( t\right) \right\rangle &
=e^{-(\gamma _{j}+\gamma _{k})t}\left\langle a_{j}\left( 0\right)
a_{k}\left( 0\right) \right\rangle  \notag \\
\left\langle a_{j}^{\dag }\left( t\right) a_{k}\left( t\right) \right\rangle
& =e^{-(\gamma _{j}+\gamma _{k})t}\left\langle a_{j}^{\dag }\left( 0\right)
a_{k}\left( 0\right) \right\rangle  \tag{B6}
\end{align}%
for different modes.

\section*{\textbf{Appendix C: Wigner functions of remotely generated states}}

The WF for the C3MSV is%
\begin{align}
W_{\rho _{123}}& =\frac{8}{\pi ^{3}}e^{-2[\left( 2\bar{n}_{1}+1\right)
\left\vert \beta _{1}\right\vert ^{2}+\left( 2\bar{n}_{2}+\allowbreak
1\right) \left\vert \beta _{2}\right\vert ^{2}+\left( 2\bar{n}_{3}+1\right)
\left\vert \beta _{3}\right\vert ^{2}]}  \notag \\
& \times e^{-8[\text{Re}(c\epsilon _{1}^{\ast }\beta _{1}\beta
_{2}\allowbreak )+\text{Re}(\epsilon _{1}^{\ast }\epsilon _{2}\beta
_{1}\beta _{3}^{\ast })+\text{Re}(c\epsilon _{2}^{\ast }\allowbreak \beta
_{2}\beta _{3})]},  \tag{C1}
\end{align}%
which has the Gaussian form.

The analytical WFs of $\rho _{B_{a}|A}$s are given as follows. For
convenience of writing, we set $\omega _{0}=c^{2}+s^{2}=\cosh 2r$, $\omega
_{1}=c^{2}-s^{2}\cos 2\phi $, and $\omega _{2}=c^{2}+s^{2}\cos 2\phi $. If $%
\phi =0$, then $\epsilon _{2}=0$, $r=r_{1}$, $\omega _{1}=1$ and $\omega
_{2}=\omega _{0}$. If $\phi =\pi /2$, then $\epsilon _{1}=0$, $r=r_{2}$, $%
\omega _{1}=\omega _{0}$, and $\omega _{2}=1$.

(a) The WF for $\rho _{1_{a}|23}$ is obtained by%
\begin{align}
W_{\rho _{1_{a}|23}}& =\frac{4e^{-2[\omega _{1}\left\vert \beta
_{2}\right\vert ^{2}+\omega _{0}\left\vert \beta _{3}\right\vert ^{2}+4\text{%
Re}(c\epsilon _{2}^{\ast }\beta _{2}\beta _{3})]/\omega _{2}}}{\pi
^{2}\omega _{2}^{3}}  \notag \\
& \times (4\left\vert c\beta _{2}+\epsilon _{2}\beta _{3}^{\ast }\right\vert
^{2}-\omega _{2}).  \tag{C2}
\end{align}%
When $\phi =0$, Eq.(C2) will be reduced to $W_{\rho _{1_{a}|23}}=(2/\pi
)e^{-2\left\vert \beta _{2}\right\vert ^{2}/\omega _{0}}(4c^{2}\omega
_{0}^{-3}\left\vert \beta _{2}\right\vert ^{2}-\omega _{0}^{-2})\times
(2/\pi )e^{-2\left\vert \beta _{3}\right\vert ^{2}}$.

(b) The WF for $\rho _{2_{a}|13}$ is obtained by%
\begin{align}
W_{\rho _{2_{a}|13}}& =\frac{4e^{-2[\omega _{1}\left\vert \beta
_{1}\right\vert ^{2}+\omega _{2}\allowbreak \left\vert \beta _{3}\right\vert
^{2}-4\text{Re}(\epsilon _{1}^{\ast }\epsilon _{2}\beta _{1}\allowbreak
\beta _{3}^{\ast })]/\omega _{0}}}{\pi ^{2}s^{2}\omega _{0}^{3}}  \notag \\
& \times (4c^{2}\left\vert \epsilon _{1}^{\ast }\beta _{1}+\epsilon
_{2}^{\ast }\beta _{3}\right\vert ^{2}-\omega _{0}s^{2}\allowbreak ).
\tag{C3}
\end{align}%
When $\phi =0$, Eq.(C3) will be reduced to $W_{\rho _{2_{a}|13}}=(2/\pi
)e^{-2\left\vert \beta _{1}\right\vert ^{2}/\omega _{0}}(4c^{2}\omega
_{0}^{-3}\cos ^{2}\phi \left\vert \beta _{1}\right\vert ^{2}-\omega
_{0}^{-2})\times (2/\pi )e^{-2\left\vert \beta _{3}\right\vert ^{2}}$; When $%
\phi =\pi /2$, Eq.(C3) will be reduced to $W_{\rho _{2_{a}|13}}=(2/\pi
)e^{-2\left\vert \beta _{3}\right\vert ^{2}/\omega _{0}}(4c^{2}\omega
_{0}^{-3}\sin ^{2}\phi \left\vert \beta _{3}\right\vert ^{2}-\omega
_{0}^{-2})\times (2/\pi )e^{-2\left\vert \beta _{1}\right\vert ^{2}}$.

(c) The WF for $\rho _{3_{a}|12}$ is obtained by%
\begin{align}
W_{\rho _{3_{a}|12}}& =\frac{4e^{-2[\omega _{2}\left\vert \beta
_{2}\right\vert ^{2}+\omega _{0}\left\vert \beta _{1}\right\vert ^{2}+4\text{%
Re}(c\epsilon _{1}^{\ast }\beta _{1}\beta _{2})]/\omega _{1}}}{\pi
^{2}\omega _{1}^{3}}  \notag \\
& \times (4\left\vert c\beta _{2}+\epsilon _{1}\beta _{1}^{\ast }\right\vert
^{2}-\omega _{1}).  \tag{C4}
\end{align}%
When $\phi =\pi /2$, Eq.(C4) will be reduced to $W_{\rho _{3_{a}|12}}=(2/\pi
)e^{-2\left\vert \beta _{2}\right\vert ^{2}/\omega _{0}}(4c^{2}\omega
_{0}^{-3}\left\vert \beta _{2}\right\vert ^{2}-\omega _{0}^{-2})\times
(2/\pi )e^{-2\left\vert \beta _{1}\right\vert ^{2}}$.

(d) The WF for $\rho _{2_{a}3_{a}|1}$ is obtained by%
\begin{align}
W_{\rho _{2_{a}3_{a}|1}}& =\frac{2e^{-2\left\vert \beta _{1}\right\vert
^{2}/\omega _{2}}}{\pi \omega _{0}\omega _{2}^{5}}  \notag \\
& \times \lbrack \omega _{2}^{2}\allowbreak \omega _{1}+4\left\vert \epsilon
_{1}\beta _{1}\right\vert ^{2}(4c^{4}+4c^{2}\left\vert \epsilon _{1}\beta
_{1}\right\vert ^{2}-\omega _{1}^{2})]  \tag{C5}
\end{align}%
When $\phi =\pi /2$, Eq.(C5) will be reduced to $W_{\rho
_{2_{a}3_{a}|1}}=(2/\pi )e^{-2\left\vert \beta _{1}\right\vert ^{2}}$.

(e) The WF for $\rho _{1_{a}3_{a}|2}$ is obtained by%
\begin{equation}
W_{\rho _{1_{a}3_{a}|2}}=\frac{2e^{-2\left\vert \beta _{2}\right\vert
^{2}/\omega _{0}}}{\pi \omega _{0}^{5}}(\omega _{0}^{2}\allowbreak
+8c^{4}\left\vert \beta _{2}\right\vert ^{4}-8\omega _{0}c^{2}\left\vert
\beta _{2}\right\vert ^{2}),  \tag{C6}
\end{equation}%
which is independent of $\phi $.

(f) The WF for $\rho _{1_{a}2_{a}|3}$ is obtained by%
\begin{align}
W_{\rho _{1_{a}2_{a}|3}}& =\frac{2e^{-2\left\vert \beta _{3}\right\vert
^{2}/\omega _{1}}}{\pi \omega _{0}\omega _{1}^{5}}\allowbreak  \notag \\
& \times \lbrack \omega _{1}^{2}\allowbreak \omega _{2}+4\left\vert \epsilon
_{2}\beta _{3}\right\vert ^{2}(4c^{4}\allowbreak +4c^{2}\left\vert \epsilon
_{2}\beta _{3}\right\vert ^{2}-\omega _{2}^{2})].  \tag{C7}
\end{align}%
When $\phi =0$, Eq.(C7) will be reduced to $W_{\rho _{1_{a}2_{a}|3}}=(2/\pi
)e^{-2\left\vert \beta _{3}\right\vert ^{2}}$.

(g) The WFs for $\rho _{1_{a}|2}$ and $\rho _{1_{a}3|2}$ are obtained by
\begin{equation}
W_{\rho _{1_{a}|2}}=W_{\rho _{1_{a}3|2}}=\frac{2e^{-2\left\vert \beta
_{2}\right\vert ^{2}/\omega _{0}}}{\pi \omega _{0}^{3}}(4c^{2}\left\vert
\beta _{2}\right\vert ^{2}-\omega _{0}),  \tag{C8}
\end{equation}%
which is independent of $\phi $.

(h) The WFs for $\rho _{3_{a}|1}$\ and $\rho _{23_{a}|1}$ are obtained by%
\begin{equation}
W_{\rho _{3_{a}|1}}=W_{\rho _{23_{a}|1}}=\frac{\allowbreak 2e^{-\allowbreak
2\left\vert \beta _{1}\right\vert ^{2}/\omega _{2}}}{\pi \omega _{2}^{3}}%
(4\left\vert \epsilon _{1}\beta _{1}\right\vert ^{2}+\omega _{2}).  \tag{C9}
\end{equation}%
When $\phi =\pi /2$, Eq.(C9) will be reduced to $W_{\rho _{3_{a}|1}}=W_{\rho
_{23_{a}|1}}=(2/\pi )e^{-2\left\vert \beta _{1}\right\vert ^{2}}$.

(i) The WFs for $\rho _{3_{a}|2}$\ and $\rho _{13_{a}|2}$ are obtained by%
\begin{equation}
W_{\rho _{3_{a}|2}}=W_{\rho _{13_{a}|2}}=\frac{2e^{-2\left\vert \beta
_{2}\right\vert ^{2}/\omega _{0}}}{\pi \omega _{0}^{3}}(4c^{2}\left\vert
\beta _{2}\right\vert ^{2}-\omega _{0}),  \tag{C10}
\end{equation}%
which is independent of $\phi $.

(j) The WFs for $\rho _{2_{a}|1}$\ and $\rho _{2_{a}3|1}$ are obtained by%
\begin{align}
W_{\rho _{2_{a}|1}}& =W_{\rho _{2_{a}3|1}}  \notag \\
& =\frac{2e^{-2\left\vert \beta _{1}\right\vert ^{2}/\omega _{2}}}{\pi
\omega _{2}^{3}}(4c^{2}\left\vert \beta _{1}\right\vert ^{2}\allowbreak \cos
^{2}\phi -\omega _{2}\cos 2\phi ).  \tag{C11}
\end{align}%
When $\phi =\pi /2$, Eq.(C11) will be reduced to $W_{\rho
_{2_{a}|1}}=W_{\rho _{2_{a}3|1}}=(2/\pi )e^{-2\left\vert \beta
_{1}\right\vert ^{2}}$.

(k) The WFs for $\rho _{1_{a}|3}$ and $\rho _{1_{a}2|3}$ are obtained by%
\begin{equation}
W_{\rho _{1_{a}|3}}=W_{\rho _{1_{a}2|3}}=\frac{2e^{-\allowbreak 2\left\vert
\beta _{3}\right\vert ^{2}/\omega _{1}}}{\pi \omega _{1}^{3}}\allowbreak
(4\left\vert \epsilon _{2}\beta _{3}\right\vert ^{2}+\omega _{1}).  \tag{C12}
\end{equation}%
When $\phi =0$, Eq.(C12) will be reduced to $W_{\rho _{1_{a}|3}}=W_{\rho
_{1_{a}2|3}}=(2/\pi )e^{-2\left\vert \beta _{3}\right\vert ^{2}}$.

(l) The WFs for $\rho _{2_{a}|3}$\ and $\rho _{12_{a}|3}$ are obtained by%
\begin{align}
W_{\rho _{2_{a}|3}}& =W_{\rho _{12_{a}|3}}  \notag \\
& =\frac{2e^{-2\left\vert \beta _{3}\right\vert ^{2}/\omega _{1}}}{\pi
\omega _{1}^{3}}(4c^{2}\left\vert \beta _{3}\right\vert ^{2}\allowbreak \sin
^{2}\phi +\omega _{1}\cos 2\phi ).  \tag{C13}
\end{align}%
When $\phi =0$, Eq.(C13) will be reduced to $W_{\rho _{2_{a}|3}}=W_{\rho
_{12_{a}|3}}=(2/\pi )e^{-2\left\vert \beta _{3}\right\vert ^{2}}$.

\end{document}